\documentclass[preprint,showpacs,prb,amsmath,amssymb]{revtex4}

\usepackage{graphicx}
\usepackage{dcolumn}
\usepackage{bm}
\usepackage{amssymb}

\begin{document}

\title{Possibility of Electron Pairing in Small Metallic Nanoparticles}

\author{Ze'ev Lindenfeld, Eli Eisenberg, and Ron Lifshitz}
\affiliation{Raymond and Beverly Sackler School of Physics and Astronomy,
Tel Aviv University, Tel Aviv 69978, Israel}

\date{\today}

\begin{abstract}
We investigate the possibility of electron pairing in small metallic nanoparticles at zero temperature. In these particles both electrons and phonons are mesoscopic, \textit{i.e.}\ modified by the nanoparticle's finite size. The electrons, the phonons, and their interaction are described within the framework of a simplified model. The effective electron-electron interaction is derived from the underlying electron-phonon interaction. The effect of both effective interaction and Coulomb interaction on the electronic spectrum is evaluated. Results are presented for aluminum, zinc and potassium nanoparticles containing a few hundred atoms. We find that a large portion of the aluminum and zinc particles exhibit modifications in their electronic spectrum due to pairing correlations, while pairing correlations are not present in the potassium particles.

\end{abstract}
\maketitle

\section{Introduction}\label{sec:introduction}

Pairing effects in metallic nanoparticles have been investigated extensively since the experimental work by Black, Ralph, and Tinkham \cite{ralph2,ralph3} in the mid 90's. Intensive theoretical studies of pairing interactions in nanometric particles followed the experimental work.\ \cite{delft1,smith,braun1,braun2,braun3,mastellone,berger,dukelsky,sierra,gladilin,tanaka1,tanaka2} These studies focused mainly on the influence of size quantization of the electronic levels on the paired state. The systems considered were relatively large (radius larger than 3nm) containing more than $10^{4}-10^{5}$ valence electrons, and having irregular shapes, such that the only symmetry of the problem is time reversal symmetry. Smaller irregular grains are expected to be unpaired, since their average energy-level spacing exceeds the bulk pairing gap, breaking the Anderson criterion.\ \cite{delft1,anderson1}

More recently, Kresin and Ovchinnikov \cite{kresin1,kresin2,kresin3} investigated the possibility of pairing in smaller metallic atomic clusters  containing tens of atoms. Such clusters are known to exhibit a shell structure in their electronic spectrum.\ \cite{heer1,heer2,knight,persson} This energy structure arises from the approximate spherical symmetry of the effective potential felt by the valence electrons in the nanoparticle. The near degeneracy of these shells may be considered as a narrow peak in the electronic density of states (DOS) which enhances both the energy gap and the critical temperature compared to the bulk material. Kresin and Ovchinnikov predicted a large enhancement in both parameters ($T_{c}$ between 100K to 200K and an energy gap of tens of meV) for specific examples of individual aluminum and gallium clusters containing a few tens of atoms. Similar results were qualitatively predicted by Friedel in 1992.\ \cite{friedel} The effect of shell structure in larger spherical nanoparticles, together with modifications in the effective interaction due to alteration of the electrons wave functions, as well as a nonuniform gap parameter, were considered in a recent study.\ \cite{croitoru} A large and strongly size dependent energy gap and critical temperature were predicted for these particles. The effect of shell structure on pairing was also discussed in the context of fermionic atoms in harmonic atomic traps.\ \cite{heiselberg}

A high DOS at the highest occupied shell (HOS) of the atomic cluster (or nanoparticle) is maintained as long as the particle retains spherical symmetry. However, clusters (and nanoparticles) with a non-full HOS undergo a Jahn-Teller deformation \cite{heer2}, which lifts the degeneracy of the HOS and reduces the high DOS. Therefore, Kresin and Ovchinnikov \cite{kresin1,kresin2,kresin3} considered clusters in which the deformation is minimal (\textit{i.e.}\ either clusters with completely full electronic shells, or with an almost full outer shell) and in which the energy difference between the HOS and the lowest unoccupied shell (LUS) is small, as the most favorable scenario for the observation of pairing correlations.

The energy gap and the critical temperature measured in ``large'' irregularly shaped nanoparticles of various superconductor metals are larger than the corresponding values of the bulk material. Evidence for this increase is found in measurements preformed with single aluminum particles.\ \cite{ralph2,ralph3} Additional support for this behavior comes from past work done with several types of superconducting thin films.\ \cite{kammerer,abeles,cohen1,cohen2,watson,zeller1,zeller2,watton,pettit} These films were composed of superconducting grains embedded on top of an oxidized substrate, and separated by the dielectric barriers formed by the oxidized substrate. Evidence for a pairing energy gap was found in tin grains as small as 2.5nm.\ \cite{zeller1,zeller2}

The increase in both energy gap and critical temperature was usually attributed to a larger effective phonon-mediated electron-electron interaction, due to ``soft'' surface phonons of the individual particles.\ \cite{watton,watson,pettit,deutscher2} There are also some indications for enhancement of the electron-phonon coupling in some molecular devices.\ \cite{Park_57,Pasupathy_203,Weig_046804,Sapmaz_026801} However, Kresin and Ovchinnikov \cite{kresin1,kresin2,kresin3} as well as Friedel \cite{friedel} assumed a constant phonon-mediated pairing interaction, which was taken to be equal to the corresponding values in the bulk material.

The increase in the energy gap in larger irregularly shaped nanoparticles was also explained as being due to finite size modifications in the bulk electronic DOS \cite{garcia,farine}, and to modifications in the bulk electronic wave functions which induce alteration in the effective interaction between the electrons.\ \cite{garcia} However, possible changes in the underlying electron-phonon interaction, which may be important in smaller nanoparticles, were ignored in these works. Furthermore, a semi-classical approximation, together with an assumption of a small ratio between the mean level spacing and the bulk gap, were used in these works. In the present work, we consider much smaller nanoparticles (containing only hundreds of atoms) for which the semi-classical approximation may not be appropriate, and the mean level spacing even within the HOS is larger than the bulk energy gap. Therefore, in what follows we propose a somewhat different approach which, although limited to simple geometries, is more appropriate for the description of pairing in nanoparticles with few hundreds of atoms.

In this work we present a calculation of the energy gap in nanoparticles at zero temperature, taking into account the finite size effects on both electrons and phonons. The nanoparticles are assumed to be isolated both electrically and mechanically from the outer environment. We treat spherical or nearly spherical particles. For this type of particles we explicitly evaluate the electron-phonon interaction, the resulting pairing interaction, and the Coulomb interaction. Thus, we are able to predict deviations from the results obtained using the average bulk values of these interactions.

Non-zero temperature and the transition between the paired and the unpaired state as temperature is varied are not addressed in this paper. We note that the transition is expected to be a ``smeared'' unlike the one found in an infinite system. Furthermore, the usual BCS ratio between the energy gap and critical temperature should not be used in order to calculate the characteristic temperature of the transition.

Three types of materials are considered as examples -- aluminum, zinc, and potassium. Aluminum is a typical low-temperature, weak-coupling superconductor with the second highest critical temperature (in the bulk) of the five superconducting metals -- Al, Ga, In, Zn, and Cd, that exhibit electronic shell structure in their atomic clusters and ultra-small nanoparticles.\ \cite{kresin2,heer2} Furthermore, aluminum nanoparticles (albeit larger than the ones we are considering) are used in experiments examining the properties of superconducting nanoparticles. Although potassium is a non-superconducting metal in the bulk its atomic clusters and nanoparticles exhibit a pronounced electronic shell structure.\ \cite{heer1,heer2} Therefore, the effect of the shell structure, together with modifications in the phonon-mediated electron-electron interaction, may enable the formation of a paired electronic ground-state. Finally, we expect zinc particles to be less susceptible to Janh-Teller deformations than the other four superconducting metals mentioned above, due to the larger shear modulus of zinc.

Pairing was proposed as a mechanism that may explain the apparent odd-even staggering in the ionization potential spectra (meaning that the ionization potential of a cluster with an even number of atoms is larger than that of the neighboring odd clusters) and in the abundance spectra of alkali clusters containing up to a few tens of atoms.\ \cite{barranco} Pairing interaction was also proposed as an explanation for some of the discrepancies between the calculated ionization spectrum (obtained by using independent particle models), and the measured spectra of sodium clusters containing tens of atoms.\ \cite{barranco,kuzmenko1} However, it was found that, at least when considering these extremely small metal clusters (with up to about 100 atoms), the strong fluctuations in the ionization potential could be explained by the deformation of open shell clusters \cite{ekardt3,ekardt4} and by the effect of spin degeneracy in the deformed clusters \cite{manninen} (where the only remaining degeneracy is the double degeneracy due to the spin) without resorting to the effects of pairing interaction. Therefore, we avoid the regime of very small atomic clusters and concentrate on nanoparticles containing hundreds of atoms.

The general scheme of our calculation is as follows:
\begin{enumerate}
\item Using a simple and crude model, the single-electron and single-phonon states and spectra are evaluated.
\item The electron-phonon interaction is calculated assuming a static Thomas-Fermi screened interaction.
\item The phonon-mediated pairing interaction is evaluated using either the Fr\"{o}hlich transformation or the similarity renormalization.
\item The splitting of the electronic degenerate levels caused by deviations from spherical symmetry of the particles is estimated.
\item As a reference, the modification of the electronic spectrum due to the pairing interaction is calculated in spherical particles.
\item The reduction in the effects of pairing on the low energy excitations of the electronic spectrum due to small deformations is evaluated.
\item Finally, the reduction of the pairing interaction due to Coulomb repulsion is also taken into account.
\end{enumerate}

We note that the effective interaction is initially calculated by applying the Fr\"{o}hlich transformation \cite{frohlich1} to the system's Hamiltonian. This method is easy to apply and it describes correctly some of the features of the pairing interaction. However, it produces unrealistic results when applied to a system with non-degenerate electrons. It therefore can be used only for the spherical particles. In order to account for the effect of deformations, we use the similarity renormalization method.\ \cite{glazek} This method is known to give accurate results for the transition temperature and the energy gap in strong coupling bulk material superconductors within the framework of a Hamiltonian theory.\ \cite{mielke1,mielke2}

The pairing Hamiltonian of the deformed particles involves varying coupling coefficients and non-uniform energy level distribution. We are unaware of an exact solution to such a Hamiltonian. Therefore, we use the simple BCS grand-canonical approximation to investigate this Hamiltonian, although, in principal, a canonical ensemble approach, such as fixed-N projection of the BCS state \cite{braun1,braun3,tinkham,ring}, is  more appropriate. Several authors \cite{janko,matveev,yuzbashyan,garcia} have pointed out that finite level spacing should modify the BCS solution, leading to a smaller lowest pairing excitation energy compared to the one obtained by the BCS approximation. This reduction is found to be approximately equal to the level spacing of the single-electron spectrum as long as the relevant level spacing (the level spacing within the HOS in our model) is smaller than the lowest pairing excitation energy. In our model we find that the lowest excitation energy is several times larger than the HOS level spacing, and therefore this finite-level-spacing correction can be neglected as long as one is merely interested in a rough estimate of the modifications to the lowest excitation energy. Therefore, for our purposes the simplicity of the BCS approximation makes it an adequate choice for calculating the excitation energy (see some additional comments on this subject in section \ref{subsec:The BCS model}). Nevertheless, a more accurate description of the pairing correlations should involve a canonical ensemble treatment.

The structure of the paper is as follows. In section \ref{sec:The electrons, the phonons, and the interaction between them} we describe the single particle states of both electrons and phonons in spherical particles, as well as the interaction between them. The derivation of the effective  electron-electron interaction and the resulting approximate model pairing Hamiltonian is given in section \ref{sec:The effective electron-electron interaction}. The deformation of the spherical particles is estimated in section \ref{sec:Small deformation of the spherical particles}, while the effect of Coulomb repulsion is discussed in section \ref{sec:The effect of the coulomb interaction}. The modification of the free electrons spectrum is addressed in section \ref{sec:Modification of the electrons spectrum}. Numerical results concerning both the degenerate and the non-degenerate scenarios are presented in section \ref{sec:Results}. Some measurable quantities, which may be used to identify pairing in such small nanoparticles, are briefly discussed in section \ref{sec:Additional possible measurable quantities}. Our main conclusions are summarized in section \ref{sec:Summary}.

\section{Electrons, Phonons, and the interaction between them}\label{sec:The electrons, the phonons, and the interaction between them}

\subsection{Single electron states}\label{subsec:The single-electron states}

The delocalized electrons of a nanoparticle are modeled as free particles within an infinite spherical potential well. Alternatively, one may use a spherically symmetric harmonic potential. The former effective potential corresponds to complete screening of the electrostatic potential of the particle's positive ionic background, while the latter reflects a complete lack of screening. As we comment later, it is reasonable to assume that the screening picture of the bulk material essentially persists even in nanoparticles that contain hundreds of atoms. Thus, we may expect that the electrostatic interaction is almost completely screened. Furthermore, the major features of the shell structure observed in the abundance and ionization spectra of atomic clusters, reflect the filling of major energy shells that are characterized by an angular quantum number $l$ and a radial quantum number $n$.\ \cite{heer1,heer2} This fits well with the spherical box approximation, although a deformed 3d oscillator gives an even better fit.\ \cite{heer1,heer2} The realistic effective potential lies somewhere between these two extremes, and includes deviations from spherical symmetry.\ \cite{heer1,heer2} Note, however, that the 3d harmonic potential introduces an artificially high degeneracy of the energy levels, which would lead to an enhancement of the pairing effects we are exploring. This additional degeneracy is lifted even by a small amount of screening. Thus, the infinite well approximation captures better the physics of the true potential.

The energy levels of electrons in a spherical box are given by
\begin{equation}\label{eq:electrons kinetic energy}
\epsilon_{ln}=\frac{\hbar^{2}k_{ln}^{2}}{2m^{*}},
\end{equation}
and their corresponding wave functions are
\begin{equation}\label{eq:electrons wave}
\psi_{lmn}=\left\{
\begin{array}{ll}
B_{ln}j_{l}(k_{ln}r)Y_{lm}(\theta,\phi) &  r\leq R\\
0 & r>R,
\end{array}
\right.
\end{equation}
where $l$ and $m$ are the usual angular momentum quantum numbers, with $l\geq 0$ and $-l\leq m \leq l$. Also, $j_{l}$ are the
spherical Bessel functions, $k_{ln}$ is the $n$th zero of the function $j_{l}$ divided by the radius $R$ of the sphere, $m^{*}$ is the effective mass
of the electrons, and $Y_{lm}$ are the spherical harmonics. We assume that the effective mass of the electrons remains the same as in the bulk
metal.\ \cite{kresin4} The normalization constant $B_{ln}$ is given by
\begin{equation}\label{eq:electrons normalization}
B_{ln}=\frac{\sqrt{2}}{R^{\frac{3}{2}}}(j_{l+1}(k_{ln}R))^{-1}.
\end{equation}
We note that, including spin degeneracy, each energy level may contain up to $4l+2$ electrons.

The electron field operator is given by
\begin{equation}\label{eq:the electrons field}
\Psi(\mathbf{r})=\sum_{lmn\sigma}\psi_{lmn}(\mathbf{r})c_{lmn\sigma},
\end{equation}
and the second quantized Hamiltonian takes the usual form
\begin{equation}\label{eq:electron hamiltonian}
H_{e}=\sum_{lmn\sigma}\epsilon_{ln}c^{\dag}_{lmn\sigma}c_{lmn\sigma},
\end{equation}
where $\sigma$ is the spin index.

\subsection{Single phonon states}\label{subsec:The single phonon states}

Acoustic phonons are described as elastic disturbances in the positive ionic background of the nanoparticles, which we model by using linear elasticity theory. Since the typical nanoparticles we are dealing with contain few hundreds of atoms, the use of an elastic description could be questioned.
However, the applicability of elasticity theory for nanoparticles is supported by a reasonable description of the low frequency acoustic modes for nanometer size particles.\ \cite{Duval,Murray1,Murray2,combe,ramirez} Also, phonon related phenomena in nanomechanical systems are often modeled by using an elastic model for the description of the vibrational degrees of freedom.\ \cite{geller1,lifshitz1,lifshitz2} Therefore, although in principle one may solve the discrete equations of motion for the few hundred atoms, we prefer the simple elastic model that enables an analytic expression for the effective interaction between the electrons.

The small size of particles under consideration leads inevitably to effective elastic constants that deviate from the bulk material values. Generally, nanometer size systems tend to exhibit smaller elastic coefficients compared to the bulk. However, ab-inito density functional theory calculations of the effective modulus of dilation of small silicon, tin, and lead atomic clusters and the effective Young modulus of ultra-thin Si nanowires \cite{shoabna1,shoabna2}, indicate that the softening of the elastic constants is quite restricted. For example, the calculated reduction of the effective modulus is about a factor of two for silicon clusters containing as few as 15 atoms, and about 25\% for tin and lead particles with the same number of atoms. The calculated variation in the Young modulus of silicon wires is even smaller. We also note that experimental studies have found that the Young modulus of gold nanorods (with diameters varying between 10nm-20nm) is either lowered by 20\%-30\% \cite{hristina,min} or remains essentially unchanged \cite{zijlstra} compared to the bulk material. We therefore use both the Lam\`{e} constants of the bulk material, as well as 25\% smaller Lam\`{e} constants for comparison.

We solve the linear elasticity equation of motion under stress-free boundary conditions imposed on the surface of the sphere. The relevant results of the elastic solution are given in appendix \ref{app:The longitudinal phonon sub-system} based on the treatment by Eringen.\ \cite{eringen} The normal modes of vibration can be divided into two types: spheroidal modes containing both longitudinal and transverse components, which are coupled through the boundary conditions, and torsional modes consisting solely of transverse components. Similar to the electronic wave functions and energy levels, the vibrational modes are characterized by two angular momentum quantum numbers $l$ and $m$, and by a radial quantum number $n$.

As will be explained in the following section, we are interested only in the longitudinal component of the spheroidal modes. This component can be written as
\begin{equation}\label{eq:u longitudinal}
\mathbf{u}^{\textrm{sph-lo}}=\frac{1}{p}\mathbf{\nabla}\phi,
\end{equation}
where $\phi$ is the solution of the scalar Helmholtz equation
\begin{equation}\label{eq:scalar helmholtz}
(\mathbf{\nabla}^{2}+p^{2})\phi=0,
\end{equation}
which for stress-free boundary conditions takes the form
\begin{equation}\label{eq:scalar potential}
\phi_{lmn}=A_{ln}j_{l}\left(p_{ln}r\right)Y_{lm}(\theta,\phi),
\end{equation}
where $p_{ln}=\omega_{ln}/c_{\textrm{lo}}$ ($c_{\textrm{lo}}$ is the bulk longitudinal sound velocity of the material). The boundary conditions impose a discrete spectrum, where the eigenfrequencies are characterized by the angular quantum number $l$ and by the radial quantum number $n$. Furthermore, the ratio between the amplitude $A_{ln}$ of the longitudinal component and the amplitude $C_{ln}$ of the transverse component are uniquely determined by the boundary conditions \cite{eringen} (see appendix \ref{app:The longitudinal phonon sub-system}). We note that a continuum approach results in an infinite number of normal modes. In order to account for the atomistic discreteness of the nanoparticles, we take into account only the lowest $3N_{a}$ modes of the entire spectrum, where $N_{a}$ is the number of atoms in the nanoparticle. We define the Debye energy of a nanoparticle  as the energy of the most energetic phonon mode.

In order to quantize the vibrational modes we write the spheroidal displacement field as
\begin{equation}\label{eq:quantization vibration}
\mathbf{u}^{sph}=\sum_{lmn}\sqrt{\frac{\hbar}{2\rho
\omega_{ln}}}\left (\mathbf{u^{*}}^{\textrm{sph}}_{lmn}b^{\dag}_{lmn}+\mathbf{u}^{\textrm{sph}}_{lmn}b_{lmn}\right ),
\end{equation}
where $\rho$ is the material density of the cluster and $b^{\dag}$ $(b)$ are the phonon creation (annihilation) operators. The second quantized
Hamiltonian is obtained by inserting the quantized displacement field into the linear elastic Hamiltonian, thus obtaining
\begin{equation}\label{eq:phonon hamiltonian}
H_{ph}=\sum_{lmn}\hbar\omega_{ln}\left(b^{\dag}_{lmn}b_{lmn}+\frac{1}{2}\right).
\end{equation}
The amplitude of the longitudinal component $A_{ln}$ is uniquely determined by the normalization condition imposed by the quantization procedure
\begin{equation}\label{eq:normalization}
\int_{V}\left|\mathbf{u}^{\textrm{sph}}_{lmn}\right|^{2}d^{3}r=1,
\end{equation}
and by the ratio $A_{ln}/C_{ln}$.

\subsection{Electron-phonon interaction}\label{subsec:The electron-phonon interactions}

We assume a screened interaction between the electrons and the density disturbance in the ionic background, \textit{i.e.}\ the phonons. This interaction is added to the free Hamiltonian composed of $H_{e}$ and $H_{ph}$. We take into account only the static electronic screening when considering the interaction between the electrons and the phonons.\ \cite{fetter,pines1,pines2} This approximation is reasonable since the maximal frequency of the phonons is two orders of magnitude smaller than the Fermi energy of the electrons. In order to keep the calculation as simple as possible we assume static electronic screening also when considering the direct Coulomb interaction between the electrons (see section \ref{sec:The effect of the coulomb interaction}). We apply a unitary transformation to the basic Hamiltonian ($H_{e}+H_{p}+H_{e-p}$) in order to account for the contribution of the phonons (or the phonon screening) to the effective interaction between the electrons, as in the original work of Fr\"{o}hlich.\ \cite{frohlich1}

We utilize the Thomas-Fermi approximation in order to describe the static screening. This approximation should still be reasonably valid for particles containing hundreds of atoms, since the typical bulk Thomas-Fermi screening length is an order of magnitude smaller than the diameter of such particles. Therefore, the screened electrostatic interaction Hamiltonian between the electrons and the phonons takes the form
\begin{equation}\label{eq:general e-p}
H_{e-p}=\int_{V}\int_{V}\rho_{e}(\mathbf{r_{1}})v_{TF}(\left |\mathbf{r_{1}}-\mathbf{r_{2}}\right |)\delta\rho_{i}(\mathbf{r_{2}})
d^{3}r_{1}d^{3}r_{2},
\end{equation}
where
\begin{eqnarray}
v_{TF}(r)&=& \frac{e^{-k_{TF}r}}{r} \nonumber \\
&=&4\pi k_{TF}\sum_{l=0}^{\infty}i_{l}\left (k_{TF}r_{<}\right )k_{l}\left (k_{TF}r_{>}\right
)\sum_{m=-l}^{l}Y_{lm}\left(\theta_{1}\phi_{1}\right )Y_{lm}^{*}\left (\theta_{2}\phi_{2}\right ), \label{eq:screened potential}
\end{eqnarray}
$\rho_{e}$ is the electronic local density, $\delta\rho_{i}$ is the local variation in the ionic background charge density due to the presence of the phonons, $k_{TF}$ is the Thomas-Fermi wave number, $i_{l}(kr)$ and $k_{l}(kr)$ are the modified spherical Bessel functions, $r_{>}=r_{1}$ and $r_{<}=r_{2}$ when $r_{1}>r_{2}$, and vice versa.

Considering small deformations, the relative change in an infinitesimal volume element is given (to first order in the displacement field) by $\mathbf{\nabla}\!\cdot\!\mathbf{u}$.\ \cite{landl} Therefore, to first order in $u$, the effect of the transverse component of the spheroidal phonons on $\delta\rho_{i}$ vanishes. Thus, we can take into account only the longitudinal component of the spheroidal modes when considering the interaction with the electrons.

The interaction Hamiltonian is therefore given by
\begin{equation}\label{eq:spesicifc e-p 1}
H_{e-p}=\int_{V}\int_{V}e^{2}Zn_{0}\Psi^{\dag}\left (\mathbf{r_{1}}\right )\Psi\left (\mathbf{r_{1}}\right )v_{TF}(\left
|\mathbf{r_{1}}-\mathbf{r_{2}}\right |)\mathbf{\nabla}\!\cdot\!\mathbf{u^{sph}}\left (\mathbf{r_{2}}\right
)d^{3}r_{1}d^{3}r_{2},
\end{equation}
where $Z$ is the number of valence electrons per atom, and $n_{0}$ is the atomic density of the cluster, which is taken to be equal to its bulk
value. Using Eqs.\ \eqref{eq:electrons wave}--\eqref{eq:electrons normalization}, \eqref{eq:u longitudinal}--\eqref{eq:quantization vibration}, and \eqref{eq:normalization} we obtain
\begin{equation}\label{eq:spesicifc e-p 2}
H_{e-p}=\sum_{LMN\sigma}M_{LMN}c^{\dag}_{l_{1}m_{1}n_{1}\sigma}c_{l_{2}m_{2}n_{2}\sigma}b_{l_{3}m_{3}n_{3}}+c.c.,
\end{equation}
where $L=\{l_{1},l_{2},l_{3}\}$, $M=\{m_{1},m_{2},m_{3}\}$, and $N=\{n_{1},n_{2},n_{3}\}$, and where the angular momentum quantum numbers satisfy
\begin{eqnarray}
l_{i} & \geq & 0, \; i=1,2,3 \label{eq: usual restriction on l} \\
-l_{i} & \leq & m_{i} \leq l_{i}, \; i=1,2,3  \label{eq:usual restriction on m} \\
\left |l_{1}-l_{2}\right | & \leq & l_{3} \leq l_{1}+l_{2} \label{eq:triangle l} \\
l_{1}+l_{2}+l_{3} &=& even\ integer \label{eq:even sum of ls} \\
m_{3} &=& m_{1}-m_{2}. \label{eq:conservation of mz}
\end{eqnarray}
The coupling coefficient $M_{LMN}$ is given by
\begin{eqnarray}
M_{LMN}&=&H_{LN}c\left(l_{1},l_{2},l_{3};m_{1},-m_{2},m_{1}-m_{2}\right)(-1)^{m_{2}}\nonumber\\
&=&C_{LN}R_{LN}\Theta_{L}c\left(l_{1},l_{2},l_{3};m_{1},-m_{2},m_{1}-m_{2}\right)(-1)^{m_{2}} \label{eq:defnition of M} \\
C_{LN}&=&\frac{8\pi ze^{2}n_{0}k_{TF}}{R^{3}}\sqrt{\frac{\hbar\omega_{l_{3}n_{3}}}{2\rho c_{lo}^{2}}}\frac{A_{l_{3}n_{3}}}{j_{l_{1}+1}
\left(k_{l_{1}n_{1}}R\right)j_{l_{2}+1}\left(k_{l_{2}n_{2}}R\right)}
\label{eq:CLN} \\
R_{LN}&=&\int^{R}_{0}dr_{1}r_{1}^{2}j_{l_{1}}\left(k_{l_{1}n_{1}}r_{1}\right)j_{l_{2}}\left(k_{l_{2}n_{2}}r_{1}\right)\left[k_{l_{3}}\left(k_{TF}r_{1}\right)
\int^{r_{1}}_{0}dr_{2}r_{2}^{2}j_{l_{3}}\left(p_{l_{3}n_{3}}r_{2}\right)i_{l_{3}}\left(k_{TF}r_{2}\right)\right.\nonumber
\\& &\left. +i_{l_{3}}\left(k_{TF}r_{1}\right)
\int^{R}_{r_{1}}dr_{2}r_{2}^{2}j_{l_{3}}\left(p_{l_{3}n_{3}}r_{2}\right)k_{l_{3}}\left(k_{TF}r_{2}\right)\right]
\label{eq:RLN}\\
\Theta_{L}&=&\sqrt{\frac{\left(2l_{1}+1\right)\left(2l_{2}+1\right)}{4\pi\left(2l_{3}+1\right)}}
c\left(l_{1},l_{2},l_{3};0,0,0\right),\label{eq:thetaLM}
\end{eqnarray}
where $c\left(l_{1},l_{2},l_{3};0,0,0\right)$ and $c\left(l_{1},l_{2},l_{3};m_{1},-m_{2},m_{1}-m_{2}\right)$ are Clebch-Gordan coefficients
(adopting the notation of Rose \cite{rose}).  The detailed derivation of $M_{LMN}$ is given in appendix \ref{app:The detailed derivation}. We can summarize the total Hamiltonian as
\begin{equation}\label{eq:total hamiltonian}
H=H_{e}+H_{ph}+H_{e-p}=H_{0}+H_{e-p}.
\end{equation}

\section{Effective electron-electron interaction}\label{sec:The effective electron-electron interaction}

\subsection{The Fr\"{o}hlich transformation}\label{subsec:The Frolhich transformation}
We apply the unitary Fr\"{o}hlich transformation \cite{frohlich1} to the Hamiltonian \eqref{eq:total hamiltonian},
\begin{equation}\label{eq:unitary transformation}
H_{s}=e^{s^{\dag}}He^{s}=H_{0}+H_{e-p}+\left[H_{0},s\right]+\frac{1}{2}\left[\left(H_{e-p}+\left[H_{0},s\right]\right),s\right]
+\frac{1}{2}\left[H_{e-p},s\right]+O\left(M_{LMN}^{3}\right),
\end{equation}
where $s$ is an anti-hermitian operator defined by
\begin{equation}\label{eq:s}
s=\sum_{LMN\sigma}\alpha_{LMN}\left(c^{\dag}_{l_{1}m_{1}n_{1}\sigma}c_{l_{2}m_{2}n_{2}\sigma}b_{l_{3}m_{3}n_{3}}+
c^{\dag}_{l_{2}m_{2}n_{2}\sigma}c_{l_{1}m_{1}n_{1}\sigma}b^{\dag}_{l_{3}m_{3}n_{3}}\right).
\end{equation}
The coefficients $\alpha_{LMN}$ are given by
\begin{equation}
\alpha_{LMN}=\frac{M_{LMN}}{\epsilon_{l_{1}n_{1}}-\epsilon_{l_{2}n_{2}}+\hbar\omega_{l_{3}n_{3}}},\label{eq:alphaN}
\end{equation}
and are chosen so as to eliminate the electron-phonon interaction up to second order in  $M_{LMN}$. Specifically, the particular choice
of $\alpha$ ensures that $H_{e-p}+\left[H_{0},s\right]=0$, and thus the lowest non-vanishing term in the transformed Hamiltonian is $\frac{1}{2}\left[H_{e-p},s\right].$

The interaction term contains a sum of operator products, most of which still involve the phonon operators. However, it also contains a purely electronic term, which represents the effective electron-electron interaction resulting from the original electron-phonon interaction (up to second order in $M_{LMN}$). Calculating explicitly this interaction term we obtain the following effective electron-electron interaction
\begin{eqnarray}
H_{e-e}=\sum_{\sigma\sigma^{'}}\sum_{\stackrel{l_{1}l_{2}l^{'}_{1}l^{'}_{2}}{\stackrel{{m_{1}m_{2}m^{'}_{2}}}{n_{1}n_{2}n^{'}_{1}n^{'}_{2}}}}
\sum_{l_{3}n_{3}}&\frac{M^{n_{1}n_{2}n_{3}}_{l_{1}l_{2}l_{3}m_{1}m_{2}m_{1}-m_{2}}
M^{n^{'}_{1}n^{'}_{2}n_{3}}_{l^{'}_{1}l^{'}_{2}l_{3}m_{1}-m_{2}+m^{'}_{2}m^{'}_{2}m_{1}-m_{2}}\hbar\omega_{l_{3}n_{3}}}
{\left(\epsilon_{l^{'}_{1}n^{'}_{1}}-\epsilon_{l^{'}_{2}n^{'}_{2}}\right)^{2}-\left(\hbar\omega_{l_{3}n_{3}}\right)^{2}}\nonumber\\
&\times c^{\dag}_{l^{'}_{2}m^{'}_{2}n^{'}_{2}\sigma^{'}}c^{\dag}_{l_{1}m_{1}n_{1}\sigma}c_{l_{2}m_{2}n_{2}\sigma}
c_{l^{'}_{1}m_{1}-m_{2}+m^{'}_{2}n^{'}_{1}\sigma^{'}}\label{eq:the general electron-electron interaction},
\end{eqnarray}
where the primed indexes stem from the operator $s$. The summation over $n_{3}$ is restricted only by the finite number of modes in each branch of the phonon spectrum (defined by the a specific value of $l_{3}$), due to the finite number of normal modes supported by the nanoparticle. We note that the numerator in (\ref{eq:the general electron-electron interaction}) is not always positive, unlike the numerator appearing in the similar result obtained for translationally invariant system.\ \cite{madelung}

Following Cooper's argument \cite{cooper}, the electrons in the paired state are paired in a manner that ``uses'' the attractive part of
$H_{e-e}$ in the ``most efficient way''. First, we note that the difference between adjacent electronic energy levels is typically larger than the maximal typical phonon energy by an order of magnitude or more. Therefore, we can expect to obtain a maximal attractive interaction between electrons belonging to the same energy level. Thus, we neglect inter-level phonon-mediated interaction. Next, we assume, as in the ordinary BCS theory, that the electrons are paired in a singlet state implying $\sigma^{'}=-\sigma$. Finally, due to the properties of the Clebch-Gordan coefficients appearing in \eqref{eq:thetaLM}, we find that, in order to ensure the negativity of the intra-level interaction, we must assume that the Cooper pairs are composed of electrons with opposite $z$ component of angular momentum. This restriction means that
\begin{eqnarray}
& &m_{1}-m_{2}+m^{'}_{2}=m \label{eq:m1} \\
& &m_{2}=-m \label{eq:m2} \\
& &m_{1}=-m^{'} \label{eq:m3}.
\end{eqnarray}
Eqs.\ \eqref{eq:m1}-\eqref{eq:m3} ensure that the numerator appearing in \eqref{eq:the general electron-electron interaction} is always positive for time-reversed pairs of electrons of the form $\{m\uparrow,-m\downarrow\}$. Since inter-level interaction is ignored, all fully occupied levels, or levels with one electron or one hole, are considered to be inert and the phonon-mediated interaction acts only within the HOS. This approximation holds as long as the temperature of the nanoparticle is much smaller than the level spacing around the Fermi level. Since this level spacing is typically of the order of 0.1eV for the particles we are considering, the approximation is reasonable up to a temperature of a few hundred Kelvins. Therefore, we are left with the effective Hamiltonian
\begin{equation}\label{eq:one shell hamiltonian}
H_{e-e}=\sum_{\sigma}\sum_{mm^{'}}\sum_{l_{3}n_{3}}-\frac{\left|H^{nnn_{3}}_{lll_{3}}\right|^{2}}{\hbar\omega_{n_{3}l_{3}}}\left[c
(l,l,l_{3};m,-m^{'},m-m^{'})\right]^{2}(-1)^{m-m^{'}}c^{\dag}_{lm^{'}n\sigma}c^{\dag}_{l-m^{'}n-\sigma}c_{l-mn-\sigma}c_{lmn\sigma},
\end{equation}
where $l$ and $n$ characterize the highest occupied electronic level.

The structure of the effective Hamiltonian in \eqref{eq:one shell hamiltonian} may have been anticipated in advance, since the coupling between time-reversed pairs is a common characteristic of the pairing phenomenon.\ \cite{koltun,anderson1} Our numerical calculations for the three materials under consideration, show that the coefficients $\alpha_{LMN}$ in \eqref{eq:alphaN} are always much smaller than 1, when one considers the interaction within a degenerate HOS. Therefore, neglecting higher-order terms in \eqref{eq:unitary transformation} is justified.

\subsection{The similarity renormalization}\label{subsec:The similarity renormalization}

In this section we introduce an alternative derivation of the interaction Hamiltonian. The above calculations work well for exactly degenerated HOS. However, in the next section we want to study the effects of the splitting of the initially degenerate electron energy levels due to static deformations of the spherical particles. We expect the splitting to reduce the effective electron-electron interaction compared to the spherical system. The effects of splitting cannot be accounted for correctly in the framework of the Fr\"{o}hlich transformation. Using \eqref{eq:the general electron-electron interaction} and introducing the energy splitting due to deformations, one obtains an artificial enhancement (and even divergence) of the effective interaction, when the energy splitting $\epsilon_{l_{1}n_{1}}-\epsilon_{l_{2}n_{2}}$ is of order $\hbar\omega_{l_{3}n_{3}}$. Note that even if no actual divergences are encountered, the second order expansion \eqref{eq:unitary transformation} is invalid if the denominators in \eqref{eq:alphaN} are smaller than the values of $M_{LMN}$.

Another deficiency of the Fr\"{o}hlich interaction \eqref{eq:the general electron-electron interaction} lies in the fact that it includes terms that represent coupling of electronic states by a virtual phonon whose energy is smaller than the energy separating the electronic states. In the usual BCS treatment these (repulsive) terms are avoided, since the interaction is assumed to be constant in $k$-space, with an artificial cutoff at the Debye energy of the material.\ \cite{tinkham} Lastly, using a Fr\"{o}hlich type interaction within the BCS formalism results in a large overestimate of the size of the energy gap and critical temperature of bulk superconductors.\ \cite{allen2}

An alternative derivation of the effective interaction relies on the application of the similarity renormalization method to the initial Hamiltonian \eqref{eq:total hamiltonian}. Although its application is more complex, it avoids the appearance of vanishing energy denominators. Furthermore, the resulting interaction is reduced between electrons with different energies, and it is automatically cut-off at the right energy scale. Also, the obtained effective interaction is always attractive. As shown by Mielke \cite{mielke1,mielke2} this approach yields the correct critical temperature and energy gap for strong-coupling bulk superconductors within the framework of the BCS model.

The derivation of the effective interaction is given in appendix \ref{app:detailed}, where we follow the treatment by Mielke.\ \cite{mielke1} We obtain the following effective electron-electron interaction between electrons belonging to the same energy shell
\begin{equation}\label{eq:interactionfullintext}
G_{mm^{'}}=\sum_{l_{3}n_{3}}\frac{-2\left|M_{mm^{'}l_{3}n_{3}\Lambda}\right|^{2}}
{\left|\varepsilon_{ml_{3}n_{3}}-\varepsilon_{m^{'}l_{3}n_{3}}\right|+\hbar\omega_{l_{3}n_{3}}}
\Theta\left(\hbar\omega_{l_{3}n_{3}}-\left|\varepsilon_{ml_{3}n_{3}}-\varepsilon_{m^{'}l_{3}n_{3}}\right|\right),
\end{equation}
where $\Theta$ is the Heaviside function and $\varepsilon_{ml_{3}n_{3}}$ are the single-electron energies in the HOS of the spherical or deformed nanoparticles. The Heaviside function in \eqref{eq:interactionfullintext} ensures  that a phonon can mediate interaction between two electron states only if its energy is larger than the energy separation between the two states. We note that \eqref{eq:interactionfullintext} and \eqref{eq:one shell hamiltonian} coincide if the deformations are neglected.

Corrections to the wave functions of both electrons and phonons due to deviation from spherical symmetry are neglected in \eqref{eq:interactionfullintext}. This is justified because it affects the numerator in \eqref{eq:interactionfullintext} only to second order in the perturbation. The effect of deformation on the phonon spectrum is minor, as deformations are small compared to particles' radii (see
section \ref{sec:Results}).

Our result differs from Mielke's interaction in the cutoff function appearing in $G_{mm^{'}}$ \eqref{eq:interactionfullintext}. The source of the difference between our cutoff function and Mielke's cutoff function is explained in appendix \ref{app:detailed}. However, our interaction and Mielke's coincide for the physical scenario considered by Mielke. This scenario consists of an effective interaction mediated by non-dispersive Einstein phonons, whose frequency is much larger than the electron energies. Our result is similar to the one obtained by H\"{u}bsch and Becker \cite{hubsch} using a different perturbative renormalization scheme.

Finally, we note that the electron energies are renormalized by the electron-phonon interaction, and that the effective interaction should depend on the renormalized electron energies. \cite{mielke1,mielke2,allen2} The renormalization is especially important in describing properties of strong-coupling bulk superconductors.\ \cite{mielke1,mielke2,allen2} This is another deficiency of the Fr\"{o}hlich interaction, which depends on the non-renormalized energies. By contrast, the energies appearing in Eq.\ \eqref{eq:interactionfullintext} are in fact the renormalized energies. However, renormalization of the single-electron energies is not important in our model since the electron-phonon interaction cannot lift the degeneracy of the electron shells in the spherical particles. The contribution of renormalization to the splitting of the HOS in the deformed particles is small, since it is of the order of $\alpha^{max}_{l=2}\left(M^{2}/\hbar\omega\right)$, where $\alpha^{max}_{l=2}<<1$ parameterizes the deviation from spherical symmetry [the definition of  $\alpha^{max}_{l=2}$ is given in Eq.\ \eqref{eq:alphal} in the following section]. Therefore, we neglect the effect of renormalization on the electron energies and instead use the non-renormalized electron energies. In appendix \ref{app:detailed} we show that the renormalization of the phonon energies due to the electron-phonon interaction can also be neglected.

The energy shift due to renormalization does depend on the shell quantum numbers $l$ and $n$ and the electron filling. Therefore, in a more detailed treatment that takes into account inter-shell effects (like the one carried out by Kresin and Ovchinnikov \cite{kresin1,kresin2,kresin3}) renormalization effects should be considered, especially when dealing with nearly degenerate HOS and LUS. In appendix \ref{app:detailed} we give the expression \eqref{eq:epsilon M} for renormalization of electron energies due to electron-phonon interaction within the HOS, which is responsible for most of the HOS shift. The expressions for the renormalization of electron energies and phonon energies are similar to the ones obtained for the bulk system by Mielke.\ \cite{mielke1}

\section{Deviation from spherical symmetry}\label{sec:Small deformation of the spherical particles}
Deviation from spherical shape leads to a splitting of the degenerate electron levels and to a decrease in the total energy in the free electrons of the nanoparticle. On the other hand, deviation from a spherically symmetric shape leads to an increase in the elastic energy of the nanoparticles. The magnitude of the deviation is determined by the balance between the increase in the elastic energy and the decrease in the electronic energy. A similar type of calculation was used by Kresin and Ovchinnikov \cite{kresin4} in order to estimate shape oscillations of aluminum atomic cluster (with 14 atoms), in which an electron from the full HOS is raised to the LUS.

The general expression for the elastic energy density is \cite{landl}
\begin{equation}\label{eq:elastic density}
\varepsilon_{ela}=\frac{1}{2}\lambda\left(\nabla\cdot \mathbf{u}\right)^{2}+\mu\sum_{i,j}\left[\frac{1}{2}\left(\frac{du_{i}}{dx_{j}}+\frac{du_{j}}{dx_{i}}\right)\right]^2,
\end{equation}
where $\lambda,\mu$ are the Lam\`{e} coefficients, and $\mathbf{u}$ is the displacement field. For simplicity, we assume that the local density in the nanoparticles remains the same as in the bulk material regardless of the shape of the particles. Therefore, we need to consider a distortion of the spherical shape for which
\begin{equation}
\nabla \cdot \mathbf{u}=0.\label{eq:divu}
\end{equation}
Choosing such a displacement field eliminates the first term in \eqref{eq:elastic density}. We also assume an axially symmetric distortion of the surface of the spherical particles. This distortion can be parameterized as follows
\begin{equation}
R\rightarrow R\left[1+\sum_{l}\alpha_{l}P_{l}(\cos\theta)\right],\label{eq:rtor}
\end{equation}
where $P_{l}(\cos\theta)$ are the Legendre polynomials. A displacement field that produces such a distortion while not changing the local density is given by \cite{kresin4}
\begin{equation}\label{eq:minimal u}
\mathbf{u}=\sum_{l}\frac{R^{2-l}}{l}\alpha_{l}\nabla\left(r^{l}P_{l}\left(\cos\theta\right)\right).
\end{equation}

We are interested only in a rough estimate of the magnitude of the deformation. Therefore, we do not consider the general displacement field \eqref{eq:minimal u}. Instead we only examine separately each value of $l$ in \eqref{eq:minimal u} and determine which multipolarity enables the largest decrease in the total energy compared to the spherical particles.

For a certain multipolarity $l$, the first order correction to the energy of a given electronic state (characterized by the quantum numbers $L$, $M$, and $N$) is \cite{kresin4}
\begin{eqnarray}
\delta\varepsilon^{l}_{LMN}&=&-2\epsilon_{LN}\alpha_{l}<LMN|P_{l}(\cos\theta)|LMN>\nonumber\\
&=&-2\epsilon_{LN}\alpha_{l}c\left(l,L,L;0,0,0\right)c\left(l,L,L;0,M,M\right).\label{eq:deltaE1}
\end{eqnarray}
Only even values of $0<l\leq 2L$ yield a nonzero $\delta\epsilon^{l}_{LMN}$. For a full shell the sum $\sum_{M}\delta\epsilon_{LMN}$ vanishes, and therefore we need to consider only non-full HOS.

For a given $l$ the additional elastic energy is
\begin{equation}\label{eq:elastic energy}
\delta E^{l}_{ela}=\int_{V}d^{3}r\varepsilon^{l}_{ela}=c_{l}\pi\alpha_{l}^{2}R^{3}\mu,
\end{equation}
where $c_{l}$ is a $l$-dependent constant that lies between 2 for $l=2$ and 4 for large $l$. We assume the deformation to be small and thus integrate the radial functions between zero and the original radius of the sphere. Consistently, we find the resulting deformation to be small, with a maximal aspect-ratio of 1.13 and an average aspect-ratio of about 1.05.
\begin{equation}
\Delta E_{l}=\sum_{M}\delta\varepsilon^{l}_{LMN}+\delta E^{l}_{ela},\label{eq:delta E}
\end{equation}
where the summation in \eqref{eq:delta E} is over all the occupied states in the HOS. The states are filled from $\left|M\right|=0$ and upwards until the highest relevant value of $\left|M\right|$.

We obtain the largest decrease in the energy of the nanoparticles $\Delta E^{max}_{l}$ and the corresponding amplitude $\alpha^{max}_{l}$, by differentiating $\Delta E_{l}$ with respect to $\alpha_{l}$ and equating the derivative to zero. The resulting expression for the amplitude is
\begin{equation}\label{eq:alphal}
\alpha^{max}_{l}=\frac{\varepsilon_{LN}}{c_{lo}\pi R^{3}\mu}c\left(l,L,L;0,0,0\right)\sum_{M}c\left(l,L,L;0,M,M\right).
\end{equation}
The maximal value of $\alpha^{max}_{l}$ is obtained for half filling of the HOS.

The largest $\left|\Delta E^{max}_{l}\right|$ is obtained for $l=2$ (quadrupole deformation). $\left|\Delta E^{max}_{l=2}\right|$ is larger by a factor of 2 to 10 than the second largest $\left|\Delta E^{max}_{l}\right|$ (for the relevant values of $L$), and larger than the sum of all other $\left|\Delta E^{max}_{2<l\leq 2L}\right|$ with their corresponding optimal $\alpha^{max}_{l}$.

We ignore the effect of surface tension which tends to decrease the deviation from the spherical shape. This effect is probably important because a large fraction of the atoms in the nanoparticles we study reside on the surface of the particles. Taking into account only the bulk elastic energy and ignoring the increase in surface tension we overestimate the size of the deformation and of the energy splitting. We also note that the deviation from spherical symmetry obtained here is not much larger than the effect of surface roughness due to the discrete atoms. However, surface roughness corresponds to a large $l$ deformation which results in a small energy correction $\delta\varepsilon^{l}_{LMN}$ compared to the quadrupole correction.

\section{Effect of Coulomb repulsion}\label{sec:The effect of the coulomb interaction}
We use the same type of screened Thomas-Fermi potential to describe the repulsive Coulomb interaction between the electrons. The Coulomb part of the Hamiltonian can be written generally as
\begin{equation}
H_{co}=\sum_{LMN\sigma\sigma^{'}}M^{co}_{LMN}c^{\dag}_{l_{4}m_{4}n_{4}\sigma}c^{\dag}_{l_{2}m_{2}n_{2}\sigma^{'}}
c_{l_{1}m_{1}n_{1}\sigma^{'}}c_{l_{3}m_{3}n_{3}\sigma},\label{eq:coulomb1}
\end{equation}
where $L=\{l_{1},l_{2},l_{3},l_{4}\}$, $M=\{m_{1},m_{2},m_{3},m_{4}\}$, $N=\{n_{1},n_{2},n_{3},n_{4}\}$. The Coulomb interaction matrix element is denoted by $M^{co}_{LMN}$
\begin{equation}
M^{co}_{LMN}=\frac{1}{2}e^{2}\int_{V}\int_{V}\psi^{*}_{l_{2}m_{2}n_{2}}\left(\mathbf{r_{1}}\right)\psi^{*}_{l_{4}m_{4}n_{4}}\left(\mathbf{r_{2}}\right)
v_{TF}(\left|\mathbf{r_{1}}-\mathbf{r_{2}}\right|)\psi_{l_{1}m_{1}n_{1}}\left(\mathbf{r_{1}}\right)\psi_{l_{3}m_{3}n_{3}}\left(\mathbf{r_{2}}\right)
d^{3}r_{1}d^{3}r_{2}.
\label{eq:coulomb coeff}
\end{equation}
We focus on the Coulomb matrix elements that are relevant for the interaction between time-reversed electron pairs within the same energy shell. We insert the expressions of the electronic wave functions \eqref{eq:electrons wave} and of the screened Thomas-Fermi interaction \eqref{eq:screened potential} into the Coulomb matrix element \eqref{eq:coulomb coeff}, and obtain
\begin{eqnarray}
M^{co}_{lnmm^{'}}&=&\frac{8\pi e^{2}k_{TF}}{R^{6}j^{4}_{l+1}\left(k_{ln}R\right)}\sum_{L^{'}=0}^{\infty}\sum_{M^{'}=-L^{'}}^{L^{'}}\int_{0}^{R}dr_{1}r^{2}_{1}j_{l}^{2}\left(k_{ln}r_{1}\right)
\nonumber\\
& &\left[k_{L^{'}}\left(k_{TF}r_{1}\right)\int^{r_{1}}_{0}dr_{2}r^{2}_{2}j_{l}^{2}\left(k_{ln}r_{2}\right)i_{L^{'}}\left(k_{TF}r_{2}\right)+i_{L^{'}}\left(k_{TF}r_{1}\right)
\int_{r_{1}}^{R}dr_{2}r^{2}_{2}j_{l}^{2}\left(k_{ln}r_{2}\right)k_{L^{'}}\left(k_{TF}r_{2}\right)\right]\nonumber\\
& &\int d\Omega_{1}Y^{*}_{lm^{'}}\left(\Omega_{1}\right)Y_{lm}\left(\Omega_{1}\right)Y_{L^{'}M^{'}}\left(\Omega_{1}\right)
\int d\Omega_{2}Y^{*}_{l-m^{'}}\left(\Omega_{2}\right)Y_{l-m}\left(\Omega_{2}\right)Y^{*}_{L^{'}M^{'}}\left(\Omega_{2}\right).\label{eq:hamlitonian coulomb explicit}
\end{eqnarray}
The angular integrals yield
\begin{eqnarray}
&&\int d\Omega_{1}Y^{*}_{lm^{'}}Y_{lm}Y_{L^{'}M^{'}}=\nonumber\\
&&(-1)^{m}(-1)^{L^{'}}\frac{2l+1}{\sqrt{4\pi (2L^{'}+1)}}c(l,l,L^{'};0,0,0)c(l,l,L^{'};m,-m^{'},-M^{'})\label{eq:int angular1}\\
&&\int d\Omega_{2}Y^{*}_{l-m^{'}}Y_{l-m}Y^{*}_{L^{'}M^{'}}=\nonumber\\
&&(-1)^{-m^{'}}\frac{2l+1}{\sqrt{4\pi (2L^{'}+1)}}c(l,l,L^{'};0,0,0)c(l,l,L^{'};m,-m^{'},-M^{'}),\label{eq:int angular2}
\end{eqnarray}
where $c(l,l,L^{'};0,0,0)=0$ unless $L^{'}$ is an even integer, which means that $(-1)^{L^{'}}=1$. Using Eqs.\ \eqref{eq:int angular1} and \eqref{eq:int angular2}, one obtains the Coulomb part of the Hamiltonian
\begin{eqnarray}
H_{co}=&N_{ln}&\sum_{\sigma}\sum_{m,m^{'}=-l}^{l}\sum_{L^{'}=0}^{2l}R_{lnL^{'}}\Theta_{lL^{'}}\left[c\left(l,l,L^{'};m,-m^{'},m-m^{'}\right)\right]^{2}
(-1)^{m-m^{'}}\nonumber\\
&\times &c^{\dag}_{lm^{'}n\sigma}c^{\dag}_{l-m^{'}n-\sigma}c_{l-mn-\sigma}c_{lmn\sigma},\label{eq:coulomb2}
\end{eqnarray}
where,
\begin{eqnarray}
N_{ln}&=&\frac{8\pi e^{2}k_{TF}}{R^{6}j^{4}_{l+1}\left(k_{ln}R\right)}\label{eq:Nln}\\
R_{lnL^{'}}&=&\int_{0}^{R}dr_{1}r_{1}^{2}j_{l}^{2}\left(k_{ln}r_{1}\right)\left[k_{L^{'}}\left(k_{TF}r_{1}\right)\int_{0}^{r_{1}}dr_{2}r_{2}^{2}
j_{l}^{2}\left(k_{ln}r_{2}\right)i_{L^{'}}\left(k_{TF}r_{2}\right)\right.\nonumber\\
& &\left.+i_{L^{'}}\left(k_{TF}r_{1}\right)\int_{r_{1}}^{R}dr_{2}r_{2}^{2}j_{l}^{2}\left(k_{ln}r_{2}\right)k_{L^{'}}\left(k_{TF}r_{2}\right)\right]
\label{eq:RlnL}\\
\Theta_{lL^{'}}&=&\frac{(2l+1)^{2}}{4\pi(2L^{'}+1)}c^{2}\left(l,l,L^{'};0,0,0\right).\label{eq:thetalL}
\end{eqnarray}

We cannot eliminate the Coulomb interaction between electrons within a degenerate or nearly-degenerate HOS by adding an additional anti-hermitian generator to the Fr\"{o}hlich generator \eqref{eq:s}, since such a generator contains vanishing or very small energy denominators. However, if we transform the Hamiltonian \eqref{eq:total hamiltonian} (where we replace $H_{e-p}$ with $H_{e-p}+H_{co}$) using the Fr\"{o}hlich generator \eqref{eq:s}, we obtain the expansion
\begin{equation}\label{eq:frolich with coulomb}
e^{s^{\dag}}He^{s}=H_{0}+H_{co}+\frac{1}{2}\left[H_{e-p},S\right]+\left[H_{co},S\right]+....
\end{equation}
The leading correction to the electron-electron interaction due to Coulomb interaction is just $H_{co}$. The commutator $\left[H_{co},S\right]$ does not renormalize the electron-electron interaction. Therefore, the next correction is of higher order in the commutator expansion, $\left[\left[H_{co},S\right],S\right]$, \textit{i.e.}\ of order $M^{2}M^{co}$.  We neglect these higher order terms and take into account only the lowest order correction, $H_{co}$. A similar result is obtained if the similarity renormalization is applied to an initial Hamiltonian containing the Coulomb term and the generator of the transformation is constructed to eliminate only the electron-phonon interaction (for more details see appendix \ref{app:detailed}). Thus, in all cases considered we simply add the screened Coulomb term to the phonon-mediated interaction.

The effect of Coulomb interaction on pairing correlations in the bulk is significantly reduced due to renormalization of the average Coulomb interaction constant.\ \cite{bogoliubov} The renormalization in the bulk stems from the large difference in the energy scale in which the phonon-mediated interaction acts (up to the Debye energy) and the one in which the screened Coulomb interaction acts (up to the plasma energy of the material which is approximately the Fermi energy), and as well as from the fact that one can reasonably well approximate the Coulomb interaction, over the entire energy range,  by a single average constant. However, in our system, only relatively narrow energy shells contribute to the renormalization of the Coulomb interaction. Furthermore, the inter-shell Coulomb interaction matrix elements are much smaller than the intra-shell matrix elements. Therefore, they should not contribute much to the renormalization of the intra-shell Coulomb interaction. We therefore claim that the relevant energy scale for the action of the Coulomb interaction in our system is the width of the split HOS and not the Fermi energy of the particles. Accordingly, we can take into account only the Coulomb interaction within the HOS, and the renormalization of the average Coulomb interaction is expected to be smaller than the renormalization in the bulk. A similar situation occurs in superconducting $\mathrm{C_{60}}$, where the Coulomb pseudopotential constant is not much smaller than the non-renormalized average Coulomb interaction constant.\ \cite{gunnarsson} By using the detailed Coulomb interaction matrix elements and not an average interaction, we avoid the necessity of estimating the renormalization within the HOS.

\section{Modification of the electronic spectrum due to pairing}\label{sec:Modification of the electrons spectrum}
We use two types of approximations to study the manner in which the highest occupied electronic level is split due to the pairing interaction at zero temperature. In the first approximation, which is applicable only to a degenerate HOS, we average the effective electron-electron interaction over the entire HOS, and obtain a single electron-electron coupling constant. This model (often referred to as the ``seniority model'') has the advantage that it is analytically solvable for a fixed number of electrons in the HOS.\ \cite{koltun,greiner} The second approximation consists of using the BCS grand-canonical approximation, in which we fix the average number of electrons to be equal to the true number of electrons in the HOS. In this approximation we can relax the requirement of a single coupling coefficient and of a degenerate HOS.

We note that the seniority model was recently used \cite{kuzmenko2} to evaluate several properties (such as temperature-dependent specific heat and magnetic susceptibility) of spherical nanoparticles, in which a constant pairing interaction extrapolated from the bulk material was assumed to act within a completely degenerate HOS. The effects of a uniform magnetic field were also considered.

\subsection{The seniority model}\label{subsec:The seniority model}
We replace the coupling coefficients of the Hamiltonian \eqref{eq:one shell hamiltonian} with a single average coupling constant
\begin{equation}
G=-\frac{1}{\left(2l+1\right)^{2}}\sum_{mm^{'}}\sum_{l_{3}n_{3}}\frac{\left|H^{nnn_{3}}_{lll_{3}}\right|^{2}}{\hbar\omega_{n_{3}l_{3}}}\left[c
(l,l,l_{3};m,-m^{'},m-m^{'})\right]^{2}.\label{eq:average G}
\end{equation}
The pairing Hamiltonian \eqref{eq:one shell hamiltonian} is replaced by
\begin{equation}\label{eq:seniority hamiltonian}
H_{e-e}=G\sum_{m^{'}}c^{\dag}_{m^{'}\uparrow}c^{\dag}_{-m^{'}\downarrow}\sum_{m}c_{-m\downarrow}c_{m\uparrow}.
\end{equation}
The energy levels of the Hamiltonian \eqref{eq:seniority hamiltonian} are given by
\begin{equation}\label{eq:seniority energies}
E(S,N)=\frac{G}{4}(N-S)(2\Omega-N-S+2),
\end{equation}
and their degeneracy is given by
\begin{equation}\label{eq:seniority degeneracy}
D(S)=\left\{
\begin{array}{ll}
1, & S=0\\
2\Omega, & S=1\\
\frac{2\Omega !}{(2\Omega-S)!S!}-\frac{2\Omega !}{(2\Omega-S+2)!(S-2)!}, &  S\geq 2,
\end{array}
\right.
\end{equation}
where $N$ is the total number of electrons in the HOS, $\Omega$ is equal to $2l+1$, and $S$ is the ``seniority number'', which counts the number of unpaired electrons in the HOS. In other words, $S$ is equal to twice (twice plus one) the number of broken Cooper pairs in the HOS if $N$ is even (odd). When the number of electrons in the HOS lies between 2 and $2l+1$, the seniority number has the following values
\begin{equation}\label{eq:seniority values}
S=\left\{
\begin{array}{ll}
0,2,\ldots ,N \;\; N\,\textrm{even} \\
1,3,\ldots ,N \;\; N\,\textrm{odd.}
\end{array}
\right.
\end{equation}
The expression \eqref{eq:seniority energies} for $E(S,N)$ can be used even if the number of electrons exceeds $\Omega$, but then $N$ should be taken as the number of holes in the HOS.\ \cite{greiner}

The energy difference between two adjacent levels with seniority numbers $S$ and $S+2$ is equal to
\begin{equation}\label{eq:delta E S to S plut two}
E(S+2,N)-E(S,N)=-G\left(\Omega-S\right).
\end{equation}
Thus, the energy levels become denser when considering higher values of $S$ (\textit{i.e.}\ higher energy levels). We consider the energy difference between the ground-state ($S=0$ or $S=1$) and the first excited state ($S=2$ or $S=3$) as the `energy gap'. The energy gap is equal to $-G\Omega$ if the number of electrons in the HOS is even, and to $-G(\Omega-1)$ if the number of electrons is odd. Note that, unlike bulk material, the size of the gap varies linearly and not exponentially with the magnitude of the coupling constant. This result remains approximately valid even when deformations are taken into account.

\subsection{The BCS model}\label{subsec:The BCS model}
The analytical results of the seniority model cannot be used for a non-degenerate HOS. Therefore, we employ the BCS approximation in order to analyze the pairing Hamiltonian in the non-degenerate HOS. The applicability of the BCS approach was discussed in section \ref{sec:introduction}. Here we add some comments about this subject.

The BCS approach was found to be reasonably successful in describing the ground-state and low-excited states for systems containing tens of interacting fermions, such as the nucleus.\ \cite{kerman,dietrich,rowe,balian,braun1} In particular, Braun and von Delft \cite{braun1} conclude (based on literature dealing with pairing in the nucleus) that the BCS approximation is adequate in order to describe, at least qualitatively, pairing correlations in ultrasmall nanoparticles. Indeed, considering the lowest excitation energy of pairing Hamiltonians in the context of nuclear pairing, the deviation between the BCS results and the ones obtained by more sophisticated treatments or the exact solution was found to be up to two-fold \cite{richardson1,dietrich}, while the typical differences, both in the context of nuclear pairing \cite{rowe,kerman} and electron-pairing in metallic nanoparticles \cite{braun2,braun3}, are usually smaller. We may conclude that following these comments, together with the discussion in section \ref{sec:introduction}, the usage of the BCS approximation is appropriate within the framework of our model, as long as we only aim at a rough description of the low excitation energies of the pairing spectrum of the nanoparticles.

We note that, in what follows, we do not assume the usual constant pairing interaction and gap parameter as in the usual BCS approximation. Instead we use the detailed interaction matrix elements, and solve for the $m$-dependent gap parameters $\Delta_{m}$.

The BCS gap equations for a given HOS, characterized by quantum numbers $l$ and $n$, are given by (see for example Greiner and Maruhn \cite{greiner})
\begin{equation}
\Delta_{m}=-\frac{1}{2}\sum_{m^{'}}\frac{G_{mm^{'}}\Delta_{m^{'}}}{\sqrt{\left(\varepsilon_{m^{'}}-\lambda\right)^{2}+\Delta_{m^{'}}^{2}}},
\label{eq:gap T0}
\end{equation}
where $\varepsilon_{m}$ are the energies of the electrons in the HOS as a function of the angular quantum number $m$, $\lambda$ is the chemical potential, and $G_{mm^{'}}$ is the pairing potential acting between the electron pairs $\{m,-m\}$ and $\{m^{'},-m^{'}\}$. The coupling coefficients $G_{mm^{'}}$ are given by
\begin{equation}\label{eq:vmmt}
G_{mm^{'}}=-\sum_{l_{3}n_{3}}\frac{2\left|H^{nn_{3}}_{ll_{3}}\right|^{2}}{\hbar\omega_{n_{3}l_{3}}}\left[c
(l,l,l_{3};m,-m^{'},m-m^{'})\right]^{2},
\end{equation}
if deformations are ignored, and by Eq.\ \eqref{eq:interactionfullintext} if they are taken into account. The set of $2l+1$ equations defined by \eqref{eq:gap T0} is solved together with the equation
\begin{equation}\label{eq:number of electrons}
N=\sum_{m^{'}}\left(1-\frac{\varepsilon_{m^{'}}-\lambda}{\sqrt{\left(\varepsilon_{m^{'}}-\lambda\right)^{2}+\Delta_{m^{'}}^{2}}}\right).
\end{equation}
Eq.\ \eqref{eq:number of electrons} is obtained by setting the expectation value of the number operator at the BCS ground-state to be equal to the total number of electrons in the HOS. We solve the BCS gap equations only for the case of an even number of electrons in the HOS, since the expectation value of the number operator in the BCS ground-state is an even number. In correspondence to the definition of the energy gap in the seniority model, we define the energy gap as twice the energy of the lowest BCS quasi-particle. The energy of the BCS quasi-particles is given by
\begin{equation}\label{eq:Em}
E_{m}=\sqrt{\Delta_{m}^{2}+\left(\varepsilon_{m}-\lambda\right)^{2}}.
\end{equation}

Considering only the phonon-mediated interaction in the degenerate case \eqref{eq:vmmt} one finds that the sum $\sum_{m^{'}}G_{mm^{'}}$ does not depend on $m$. Thus, $\Delta_{m}$ obtained by solving Eq.\ \eqref{eq:gap T0} is independent of $m$ (although the $G_{mm^{'}}$ are not equal to each other) and all BCS quasi-particles have the same energy. In this particular case, the energy gap obtained by the seniority model (for an even number of electrons) is reproduced by the BCS model. The gap parameter $\Delta_{m}$ becomes truly $m$ dependent only for non-average $m$-dependent interaction matrix elements, and non-degenerate HOS.

By considering deformation together with the Coulomb interaction, we obtain for a large portion of the particles an overall repulsive average interaction. In such cases the standard mean-field approximation of the gap parameters, $\Delta_{m}=\Delta$, yields an unpaired ground-state. However, we do not apply this approximation and therefore are able, in principal, to find non-trivial solutions to the gap equations. In fact, even for an entirely repulsive interaction such solutions may exist as long as the interaction is not constant.

The scenario of pairing in the presence of a repulsive interaction was first considered by Tolmachov \cite{bogoliubov}, and since then in numerous variations by many authors.\ \cite{mila,weger,lal,fibich} Furthermore, Mila and Abrahams \cite{mila} have shown that, for a bulk superconductor, a solution to the gap equation which is an odd function of $k-k_{F}$ is possible for an arbitrarily strong short-range repulsive interaction, and that this solution has a lower energy than the unpaired state. For the aluminum and zinc nanoparticles we find that at least a small part of the overall interaction matrix elements is in fact negative.

The energy of the paired ground-state (relative to the new chemical potential $\lambda$) is
\begin{equation}\label{eq:eEsup}
E_{pair}=\sum_{m}\left(\varepsilon_{m}-\lambda\right)\left(1-\frac{\varepsilon_{m}-\lambda}{\sqrt{\Delta^{2}_{m}
+\left(\varepsilon_{m}-\lambda\right)^{2}}}\right)-\sum_{m}\frac{\Delta_{m}}{2}\sqrt{1-\frac{\left(\varepsilon_{m}-\lambda\right)^{2}}
{\Delta_{m}^{2}+\left(\epsilon_{m}-\lambda\right)^{2}}},
\end{equation}
and the condensation energy of the nanoparticles is defined as the difference between $E_{pair}$ and the energy of the free-electron ground-state. The gap equations \eqref{eq:gap T0} may possess several solutions. Of those solutions we choose the one with the lowest condensation energy. However, since the different gap parameters $\Delta_{m}$ may have different signs, it is possible that the condensation energy of this solution is positive. Only if the condensation energy of the solution with the lowest pairing energy \eqref{eq:eEsup} is negative we can claim that the paired state is stable and that it represents the true ground-state of the nanoparticle. We find such solutions for a large portion of the aluminum and zinc nanoparticles with non-degenerate HOS (see section \ref{subsec:The energy gap}).

\section{Results}\label{sec:Results}

\subsection{Main results}\label{subsec:Main results}

We consider aluminum nanoparticles containing 100 to 400 atoms (or 300 to 1200 free electrons), zinc nanoparticles containing 200 to 500 atoms (or 400 to 1000 free electrons) and potassium nanoparticles with 100 to 500 atoms. Of the three metals considered, our results indicate that aluminum is the best candidate for observing pairing effects in ultra-small nanoparticles containing a few hundreds of atoms.

The average electron-electron interaction in aluminum nanoparticles is considered in section \ref{subsec:The electron-electron interaction -- aluminum}, while the average interaction in zinc and potassium nanoparticles is discussed in section \ref{subsec:The average electron-electron interaction -- zinc and potassium}. Generally, we find that the average phonon-mediated interaction in spherical aluminum and potassium nanoparticles is larger, on average , by a few tens of percent than the average phonon-mediated interaction strength extrapolated from the bulk value. For zinc particles we find that on average there is almost no deviation from the extrapolation from the average bulk phonon-mediated interaction strength. The average phonon-mediated interaction in deformed nanoparticles is on average reduced by more than 55\% compared to the average interaction in spherical particles. The deformations are more effective in splitting the HOS and reducing the average interaction in smaller particles than in larger ones.

The addition of Coulomb interaction to the phonon-mediated effective interaction results in an overall average attractive interaction in almost all spherical aluminum and zinc nanoparticles, while for spherical potassium particles it results in an overall average repulsive interaction. Most aluminum and zinc particles exhibit an overall average repulsive interaction when deformations are taken into account together with Coulomb interaction.

The energy gap in the three types of nanoparticles is discussed in section \ref{subsec:The energy gap}. Spherical aluminum particles exhibit an average energy gap of about 0.13eV when Coulomb interaction is ignored. When deformations and Coulomb interaction are taken into account, the average energy gap (for all particles for which we were able to find a solution to the gap equations with negative condensation energy) is reduced to about 0.025eV. The effect of deformations is most pronounced near half-shell filling. The average energy gap in the spherical zinc particles without Coulomb interaction is about 0.08eV, while the inclusion of both Coulomb repulsion and deformations results in an average gap of 0.007eV. The gap completely vanishes in potassium particles when deformations and Coulomb interaction are taken into account.

Although the average phonon-mediated interaction, when deformations are taken into account, is weaker in our system than the bulk extrapolation values, the resulting energy gap is still larger than the one found in bulk material because of the single-electron energy shell structure. When considering aluminum particles (with both deformations and Coulomb interaction taken into account) the average energy gap is larger by two orders of magnitude than the energy gap in bulk aluminum. Also, the energy gap obtained in aluminum particles (when deformations and Coulomb interaction are taken into account) is typically four times larger than the lowest single-electron excitation energy (LSEEE) of the aluminum particles and an order of magnitude smaller than the average HOS-LUS spacing. Therefore, the energy gap lies in an intermediate energy scale between the LSEEE and the inter-shell energy spacing.

\subsection{Average electron-electron interaction in aluminum particles}\label{subsec:The electron-electron interaction -- aluminum}

\begin{figure}
 \includegraphics[width=1.0\textwidth]{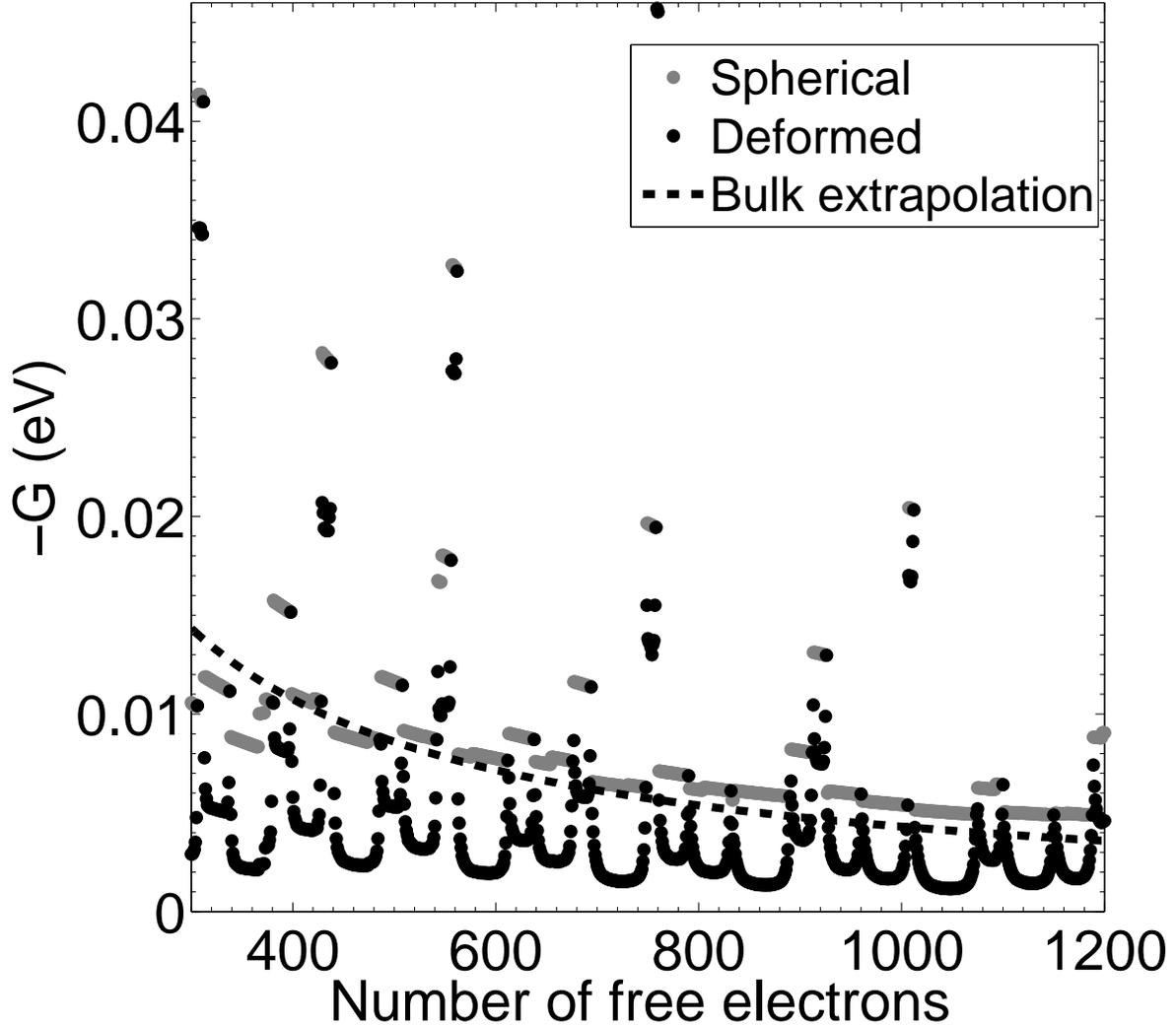}
 \caption{\label{fig1}
  Absolute value of the averaged phonon-mediated interaction constants of aluminum nanoparticles plotted as a function of the number of free electrons in the nanoparticle. The results for spherical and deformed particles are shown. The effect of Coulomb interaction is ignored. The dashed curve shows an extrapolation from the average phonon-mediated interaction obtained from the experimentally measured pairing interaction strength in the bulk, and taking the dimensionless average renormalized Coulomb interaction $\mu{*}$ to be equal to 0.1.\ \cite{anderson2}}
\end{figure}

The relevant material properties of aluminum, zinc and potassium nanoparticles are summarized in table \ref{properties}. In Fig.\ \ref{fig1} we plot the absolute value of the average effective interaction coupling constant $G$ for aluminum nanoparticles, while ignoring Coulomb interaction. The coupling constants are plotted for spherical particles and for deformed particles. The inclusion of deformations leads to an average reduction in the average coupling constants of about 56\%.
\begin{table}
 \begin{center}
 \begin{tabular}{|c|c|c|c|}\hline\hline
             & Aluminum                       & Zinc                          & Potassium   \\ \hline\hline
 $\lambda$ (GPa)\cite{simmons} \hspace{1mm}    & $\hspace{5mm}68\hspace{5mm} $ & $\hspace{5mm}46\hspace{5mm} $ & $\hspace{5mm}2.8\hspace{5mm}$ \\
 $\mu$ (GPa)\cite{simmons} \hspace{1mm}    & $\hspace{5mm}29\hspace{5mm} $ & $\hspace{5mm}51\hspace{5mm} $ & $\hspace{5mm}1.3\hspace{5mm}$ \\
 $\rho\mathrm{ (gr\cdot cm^{-3})}$\cite{simmons} \hspace{1mm}   & $\hspace{5mm}2.7\hspace{5mm} $ & $\hspace{5mm}7.3\hspace{5mm} $ & $\hspace{5mm}0.9\hspace{5mm}$ \\
 $Z$   & $\hspace{5mm}3\hspace{5mm} $ & $\hspace{5mm}2\hspace{5mm} $ & $\hspace{5mm}1\hspace{5mm}$ \\
 $\lambda_{TF} (\mathrm{nm})$\cite{ashcroft}   & $\hspace{5mm}0.049\hspace{5mm} $ & $\hspace{5mm}0.051\hspace{5mm} $ & $\hspace{5mm}0.075\hspace{5mm}$ \\
 $\hbar \omega_{D}$ (eV)   & $\hspace{5mm}0.033\hspace{5mm} $ & $\hspace{5mm}0.028\hspace{5mm} $ & $\hspace{5mm}0.008\hspace{5mm}$ \\
 $\Delta E_{\mathrm{LUS-HOS}}$ (eV) & $0.26$ & $0.33$ & $\hspace{5mm}0.11\hspace{5mm}$ \\  \hline\hline
 \end{tabular}
 \end{center}
 \begin{center}
 \begin{minipage}{0.7\textwidth}
 \caption{\label{properties}
 Basic properties of the three types of nanoparticles considered in this work. $\lambda$ and $\mu$ are the bulk material Lam\'{e} constants, $\rho$ is the bulk density, $Z$ is the number of valence electrons per atom, $\lambda_{TF}$ is the Thomas-Fermi screening length, $\hbar \omega_{D}$ is the average Debye energy of the nanoparticles, and $\Delta E_{\mathrm{LUS-HOS}}$ is the mean energy difference between the HOS and the LUS. The elastic constants and density of the aluminum, potassium, and zinc particles correspond to $T=0$, $T=4K$, and $T=4.2K$, respectively.}
 \end{minipage}
 \end{center}
\end{table}

\begin{figure}
 \includegraphics[width=1.0\textwidth]{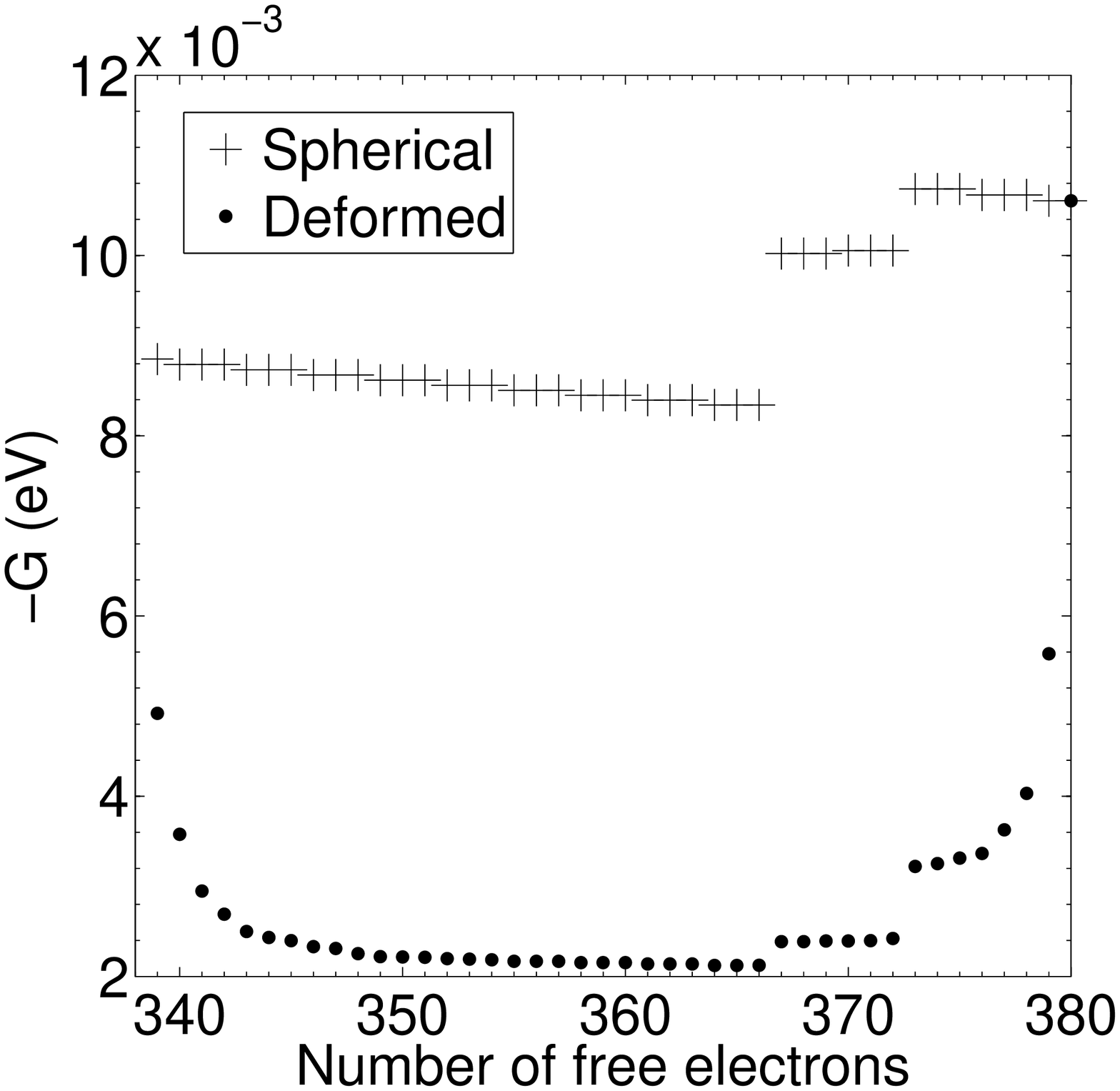}
 \caption{\label{fig2}
 Absolute value of the average phonon-mediated interaction constants $-G$ of aluminum nanoparticles containing between 339 to 380 free electrons. All particles belong to the filling of the HOS characterized by $l=10$ and $n=1$. The jumps in $-G$ between $N=366$ and $N=367$, and between $N=372$ and $N=373$, are due to the addition of a single phonon to the effective interaction.}
\end{figure}

The shell structure is clearly seen in Fig.\ \ref{fig1}. The coupling constant changes abruptly when moving from one shell to the next. Small jumps occur in the coupling constant within a certain HOS, such as when the number of free electrons in the aluminum nanoparticles is varied from $N_{e}=366$ to $N_{e}=367$ or from $N_{e}=373$ to $N_{e}=374$. These jumps arise from an increase in the number of phonon modes in the specific phonon branches that mediate the interaction between the electrons in the HOS, due to the increased number of atoms in the particle. Although these jumps may exist, their location and size should not be inferred from our calculation, due to the approximate nature of our treatment of the phonon spectrum and wave functions. The variation of the average coupling constant when a shell is filled up, as well as the small jumps due to the additional interacting phonon, are seen more clearly in Fig.\ \ref{fig2} where we plot $-G$ for the HOS characterized by $l=10$ and $n=1$.

The effect of deformation is smallest near the opening or closing of a shell. It is more or less constant at the intermediate range, especially for the smaller particles. In this intermediate range, the energy difference between most electronic levels is larger than the energy of the most energetic phonon that can mediate the interaction between the electrons. Thus, most matrix elements are in fact equal to zero, and the change in the average coupling constant between sequential nanoparticles is small. A more pronounced minimum at the HOS half-filling is observed for the larger particles for which the magnitude of the deformation is smaller.

\begin{figure}
 \includegraphics[width=1.0\textwidth]{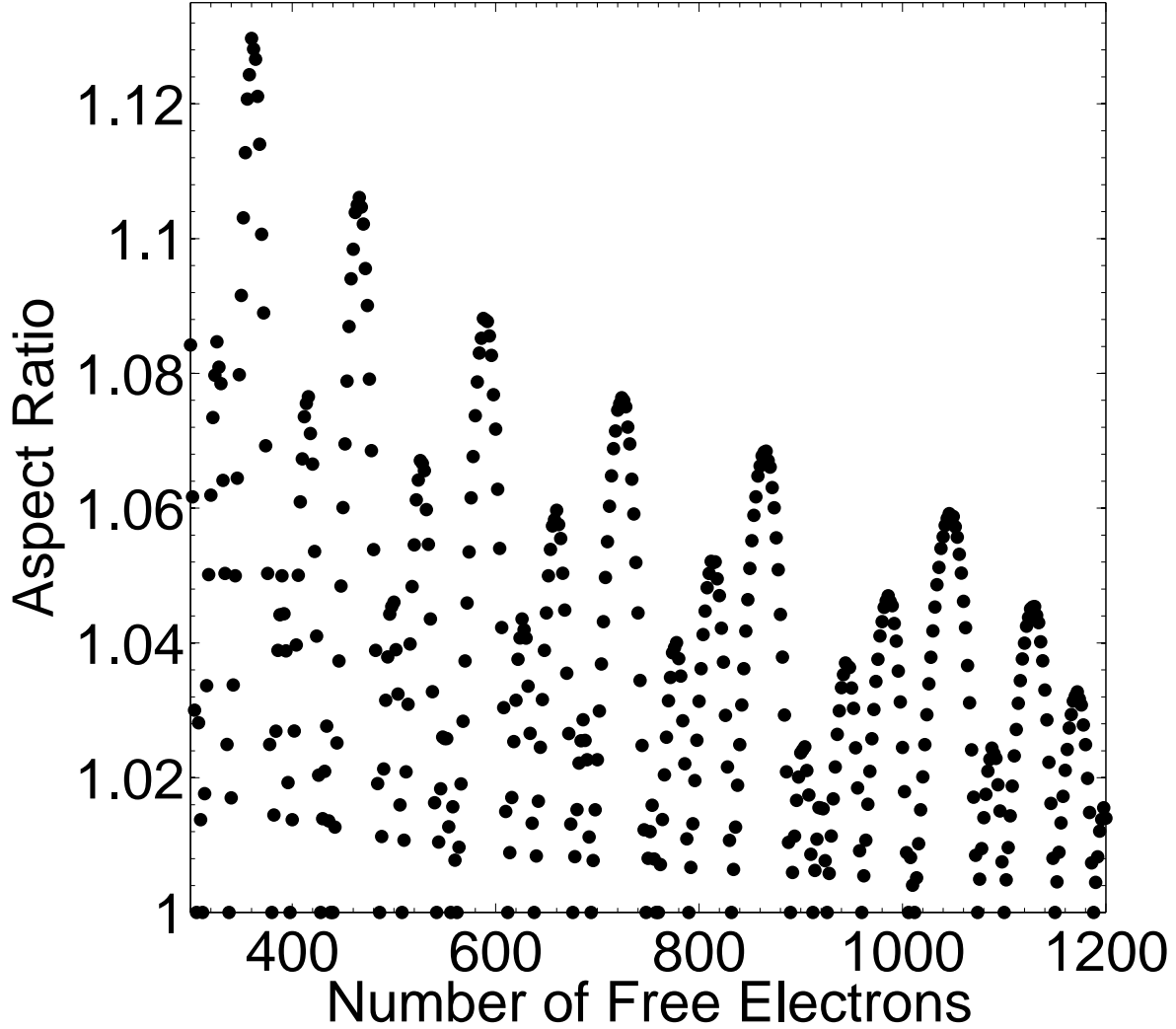}
 \caption{\label{fig3}
 Aspect ratio of deformed aluminum particles obtained using bulk aluminum elastic constants.}
\end{figure}

In Fig.\ \ref{fig3} we plot the aspect ratios of the deformed aluminum particles with even numbers of electrons in the HOS and with bulk material elastic constants. The largest deformations are obtained for the smallest particles considered, with a maximal aspect-ratio of 1.13.  The average aspect-ratio for particles with unfilled shells is 1.04.  The aspect ratio is calculated using the following expression that determines the relation between the amplitude of the quadrupole deformation $\alpha^{max}_{l=2}$ [Eq.\ \eqref{eq:alphal}] and the aspect-ratio \cite{kresin4}
\begin{equation}
\frac{c}{a}=\frac{2+2\alpha^{max}_{l=2}}{2-\alpha^{max}_{l=2}}.\label{eq:cdiva}
\end{equation}
where $c$ is the length of the major axis and $a$ is the length of the minor axis.

We compare our results to an extrapolation from the bulk coupling constant. The phonon-mediated interaction coupling constant in bulk material (denoted by $G^{ph}_{b}$) is given by \cite{tinkham}
\begin{equation}\label{eq:lambda}
G^{ph}_{b}=\frac{4\left(\lambda^{*}+\mu^{*}\right)E_{F}}{3N_{e}},
\end{equation}
where $E_{f}$ is the Fermi energy of the bulk material, $\lambda^{*}$ is the dimensionless average pairing interaction strength in the bulk material, and $\mu^{*}$ is dimensionless average renormalized Coulomb interaction (the dimensionless phonon-mediated interaction in the bulk is usually denoted by  $\lambda$). We use the value of $\lambda^{*}$ that is extracted from measurements of the critical temperature and the energy gap in bulk aluminum, and for comparison the value of $\lambda^{*}$ calculated by Morel and Anderson \cite{anderson2}. In order to obtain $G^{ph}_{b}$ we use the value $\mu^{*}=0.1$ calculated by Morel and Anderson \cite{anderson2} for aluminum. The calculation of the phonon-mediated interaction by Morel and Anderson is similar in some basic assumptions to ours since they used a free electron model, a screened Thomas-Fermi interaction between the electrons and the phonons, and they took into account only the effect of longitudinal phonons. However, unlike our model, the phonon spectrum in the Morel and Anderson calculation was approximated by an Einstein phonon model smeared into a Lorentzian line shape.

We fit our results to a function of the form
\begin{equation}\label{eq:fit}
G_{fit}=A\left(\frac{N_{e}}{750}\right)^{-\alpha},
\end{equation}
and list the values of $A$ and $\alpha$ in table \ref{fit to G}. The number 750 is just a typical value for the number of electrons considered. The values of $A$ for $G^{ph}_{b}$ [calculated using Eq.\ \eqref{eq:lambda}] are also given in table \ref{fit to G}. As can be seen from Eq.\ \eqref{eq:lambda}, $\alpha$ is identically equal to 1 for bulk material. We find that $G$ in the spherical nanoparticles is on average larger by about 40\% than $G^{ph}_{b}$ extrapolated from the experimental value in the bulk, and by about 15\% than $G^{ph}_{b}$ extrapolated from the calculations of Morel and Anderson.\ \cite{anderson2} When deformations are considered $G$ is smaller than $G^{ph}_{b}$ by about 35\% on average.

Although on average there is no large difference between the smooth extrapolation from the bulk interaction and our results, the electronic shell structure causes a large variation in the values of $G$ compared to the monotonic behavior of the extrapolation (Fig.\ \ref{fig1}). The differences between the values of $G$ in the various HOS reflects the differences in the electron-phonon matrix elements and the change in the number of phonons that can mediate the interaction in a specific HOS. The variation is especially evident when going from a HOS with high value of $l$ and low value of $n$ to a HOS with a low value of $l$ and high value of $n$ (or vise versa). For shells with small $l$ we find especially high values of $G$. However, the low degeneracy of these shells tends to cancel out the high value of $G$ when the energy gap is calculated [Eq.\ \eqref{eq:delta E S to S plut two}].

There is also a large scatter of the detailed matrix elements $G_{mm^{'}}$ for a specific particle around the average $G$ of the particle. The large scatter reflects the variation in the electron-phonon interaction when considering different $m$'s within a given shell, and especially the different number of phonons that contribute to different $G_{mm^{'}}$'s. The variation in the detailed matrix elements increases when deformations are taken into account.

In a recent work by Croitoru et al.\ \cite{croitoru} the changes in intra-shell pairing interaction in spherical nanoparticles were calculated assuming a uniform interaction, modified from its bulk average value only because of the modifications in the electronic wave functions [which were taken to be the same as the ones given in equation \eqref{eq:electrons wave}]. Croitoru et al.\ \cite{croitoru} found a large enhancement (compared to the bulk extrapolated value) in the intra-shell interaction. However, as discussed above, the effective electron-electron interaction within the HOS is anisotropic even within a degenerate HOS, where a large difference between $m$ and $m^{'}$ results in a reduced $G_{mm^{'}}$ compared to the diagonal matrix elements. The net effect is only a moderate increase in the average effective interaction, especially when the HOS is characterized by a large value of $l$. Therefore, we do not expect to find a major enhancement in the intra-shell phonon-mediated interaction, especially in large nanoparticles as the ones considered by Croitoru et al.\ \cite{croitoru}

\begin{table}
 \begin{center}
 \begin{tabular}{|c|c|c|}\hline\hline
             & $\alpha$              & A   \\ \hline\hline
 Spherical - 100\% bulk constants     & $\hspace{5mm}0.719\hspace{5mm} $ & $\hspace{5mm}0.0075\hspace{5mm}$ \\
 Deformation - 100\% bulk constants & $0.757$ & $0.0030$ \\
 Bulk - experimental & $1.0$ & $0.0057$  \\
 Bulk - Morel and Anderson \cite{anderson2} & $1.0$ & $0.0068$  \\  \hline\hline

 \end{tabular}
 \end{center}
 \begin{center}
 \begin{minipage}{0.7\textwidth}
 \caption{\label{fit to G}
 Fitting parameters for the average phonon-mediated interaction coupling constants, $G_{fit}=A\left(\frac{N_{e}}{750}\right)^{-\alpha}$, of spherical and deformed aluminum particles with bulk elastic constants, together with the corresponding approximate parameters for bulk aluminum. }
 \end{minipage}
 \end{center}
\end{table}

\begin{figure}
 \includegraphics[width=1.0\textwidth]{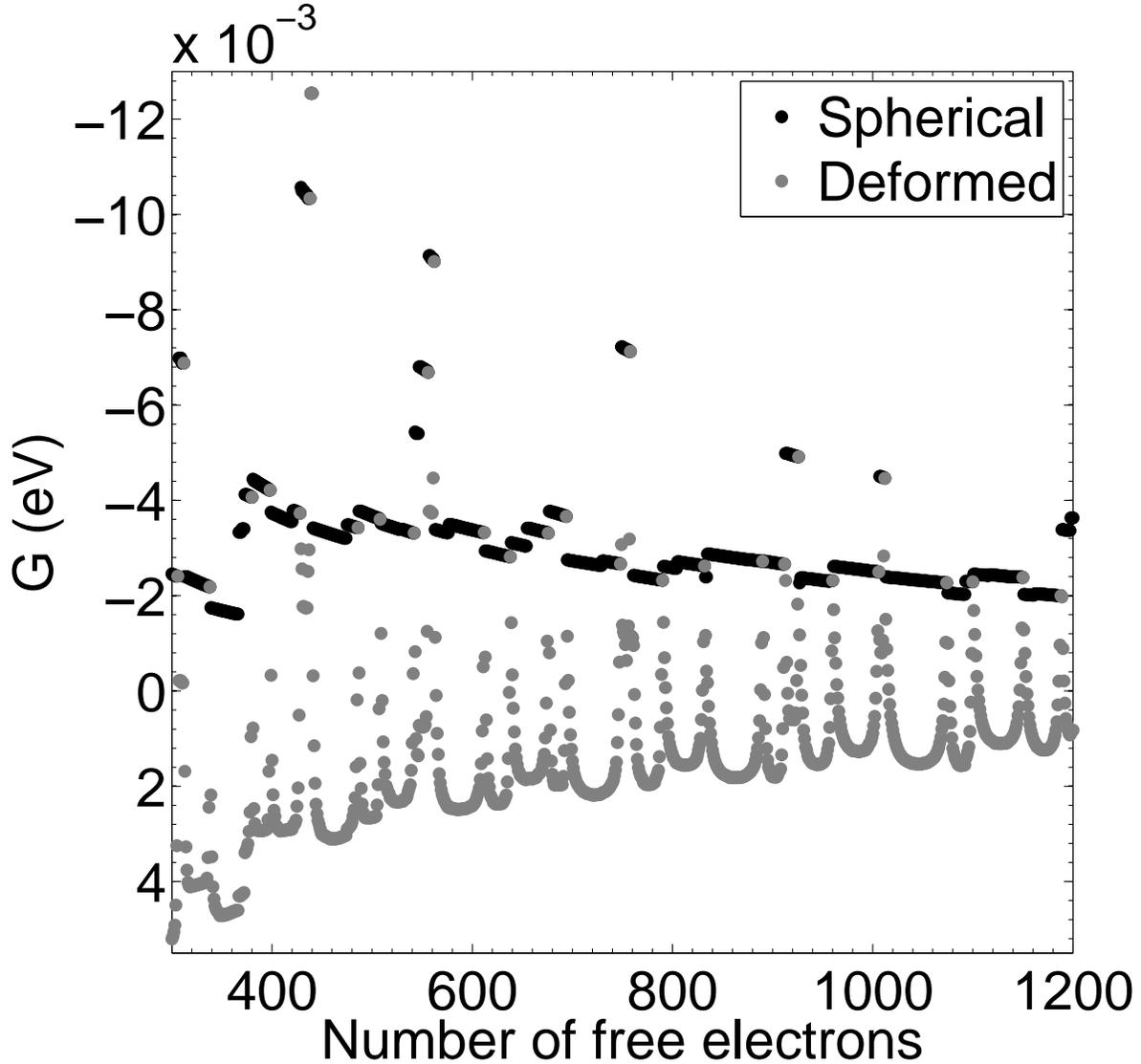}
 \caption{\label{fig4}
 Average electron-electron interaction of aluminum nanoparticles as a function of the number of free electrons in the nanoparticle, taking into account the effect of the Coulomb interaction. Results are shown for spherical and deformed particles.}
\end{figure}

Taking Coulomb repulsion into account leads to an overall average attractive interaction in the spherical particles as shown in Fig.\ \ref{fig4}. On the other hand, the combination of deformations and Coulomb interaction results in a repulsive average interaction in about 85\% of the particles. Both phonon mediated interaction and Coulomb interaction decay with the increase in the  number of electrons in the particles, but the Coulomb interaction diminishes faster. Also, the size of the deviation from spherical symmetry decreases with the increase in the size of the nanoparticles. Therefore, we find more nanoparticles with attractive average coupling as the size of the particles increases.

We also evaluate the average effective interaction and energy gap using Lam\'{e} coefficients which are 25\% smaller than the corresponding coefficients of bulk aluminum. Using these coefficients, the ratio between the longitudinal and transverse speeds of sound remain unchanged, and all dimensionless properties of the vibrational modes remain unaltered. Therefore, the electron-phonon matrix elements are changed only because of the factor $c^{-1}_{lo}=\sqrt{\rho /\left(\lambda+2\mu\right)}$ appearing in Eq.\ \eqref{eq:CLN}. The effective electron-electron interaction is proportional to the square of the electron-phonon matrix elements, and therefore it is just increased by a factor of $4/3$ in the spherical particles.

\begin{figure}
 \includegraphics[width=1.0\textwidth]{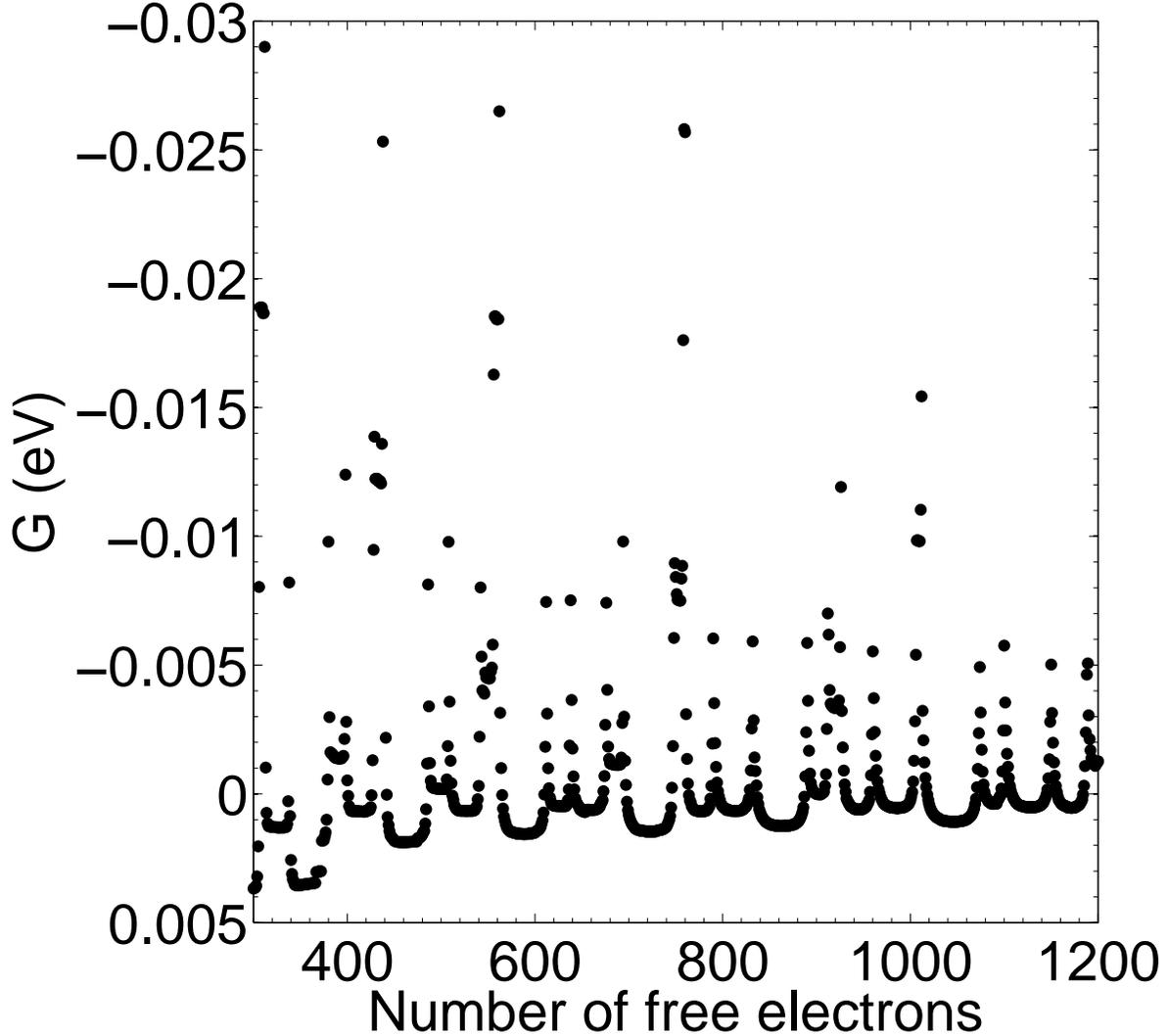}
 \caption{\label{fig5}
  Average coupling constants of deformed aluminum particles with Lam\'{e} coefficients at 75\% of their bulk value, with Coulomb interaction taken into account.}
\end{figure}

On the other hand, by decreasing the elastic constants we make the nanoparticles more susceptible to deformations. The amplitude of the deformation and therefore the size of the energy splitting, is increased by a factor of $\sqrt{64/27}$ due to the smaller Lam\'{e} coefficients [Eq.\ \eqref{eq:alphal}]. Also, the phonon frequencies decrease compared to the particles with bulk elastic constants. The lower frequencies together with the larger energy splitting decrease the number of phonons that can mediate the interaction between specific electron pairs. The net effect is nevertheless an increase of the overall electron-electron interaction, as can be seen from Fig.\ \ref{fig5} and from the fact that a larger portion of the particles examined (31\% compared to 15\%) exhibit an overall average attractive interaction when both deformations and Coulomb interaction are taken into account. We note that the ratio between the value of $G$ at the closing of a HOS and its value at half filling is larger for particles with reduced Lam\'{e} coefficients than it is for particles with bulk aluminum constants.

\subsection{Average electron-electron interaction of zinc and potassium particles}\label{subsec:The average electron-electron interaction -- zinc and potassium}

\begin{figure}
 \includegraphics[width=1.0\textwidth]{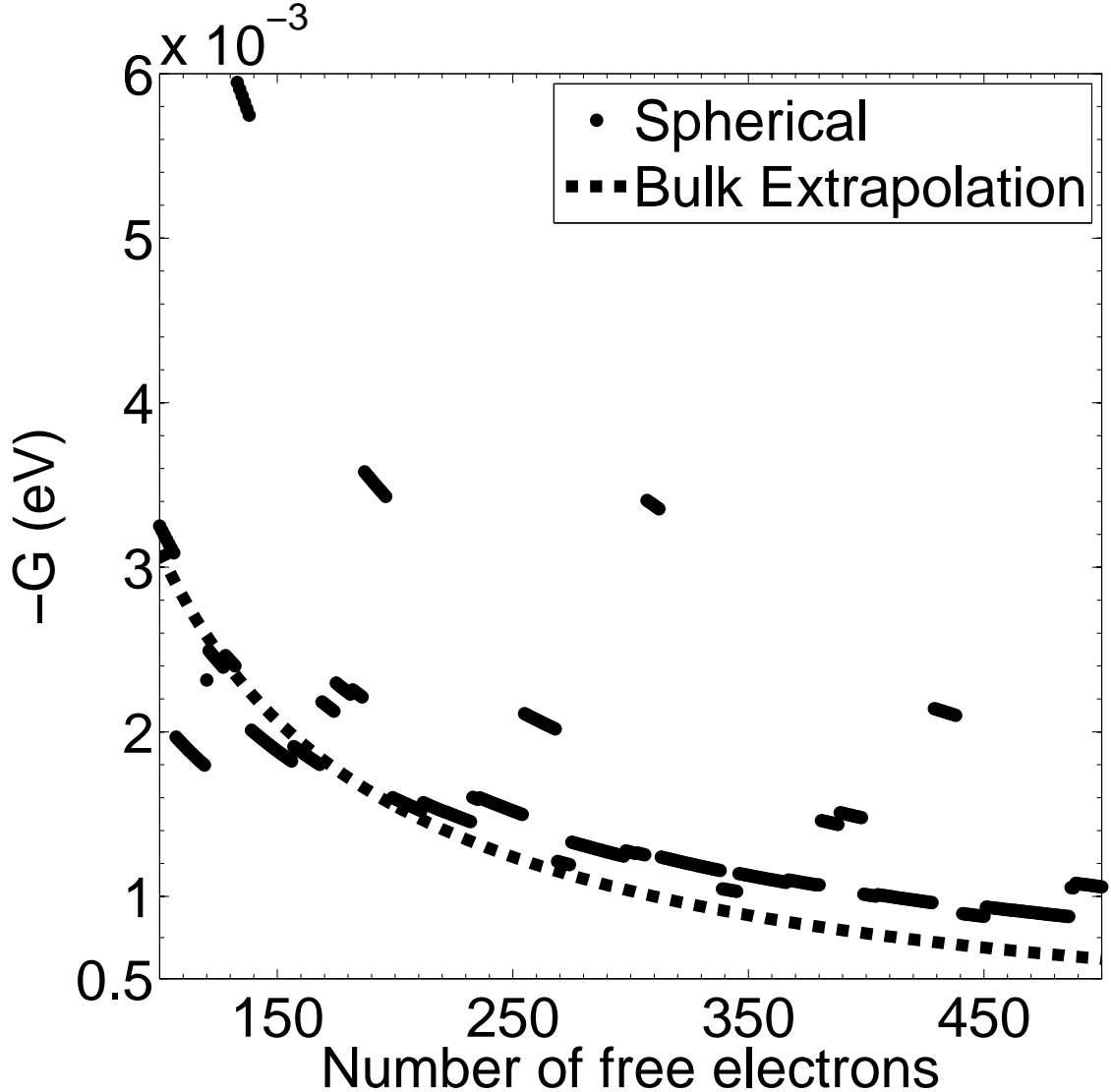}
 \caption{\label{fig6}
 Average coupling constant of spherical potassium nanoparticles without Coulomb repulsion. The extrapolation from the theoretical value of $G^{ph}_{b}$ which corresponds to $\lambda=0.11$ is also plotted.}
\end{figure}

The average coupling constants of potassium particles containing 100 to 500 atoms are shown in Fig.\ \ref{fig6}. Since potassium is a non-superconducting metal we can compare our results only to theoretical calculations of the strength of the phonon-mediated interaction. The calculations of Morel and Anderson yield a major overestimate of the overall dimensionless interaction strength. Therefore, we use the results of more detailed calculations \cite{allen1,rajput} which yield values of $\lambda$ (the dimensionless average electron-phonon interaction strength) between 0.11 and 0.16, with most results tending toward the lower values. In Fig.\ \ref{fig6} we plot the values of $G$ for potassium particles together with the calculated average phonon mediated interaction  $G^{ph}_{b}$ given in Eq.\ \eqref{eq:lambda}, using $\lambda=0.11$. The average of the ratio $G/G^{ph}_{b}$ is about 1.4 for $\lambda=0.11$ and 1.05 for $\lambda=0.16$. Unlike the aluminum particles, the addition of Coulomb interaction together with deformations results in purely repulsive interaction matrix elements. Even when deformation is disregarded, the addition of Coulomb interaction results in an overall average repulsive interaction for all potassium nanoparticles.

\begin{figure}
 \includegraphics[width=1.0\textwidth]{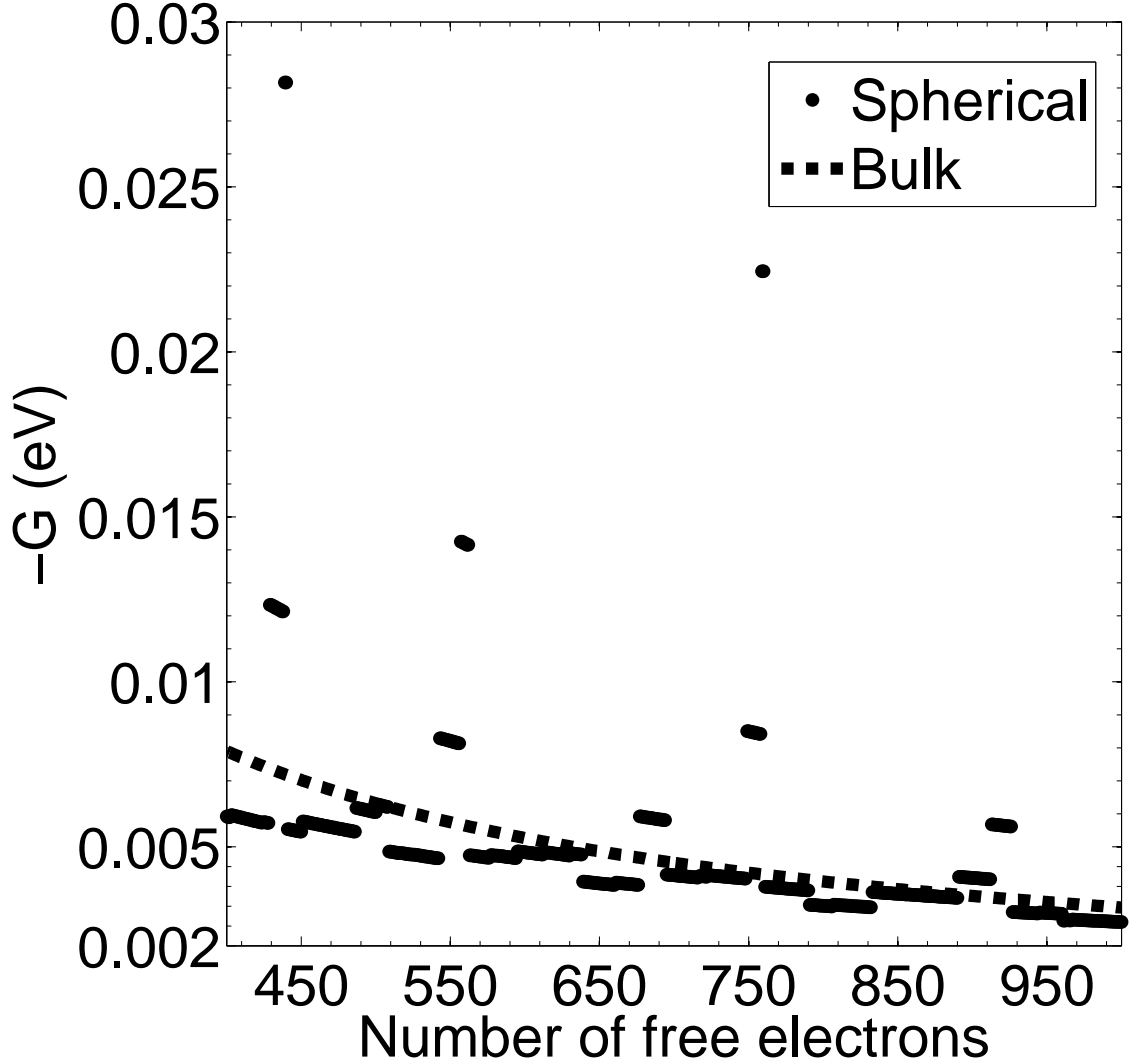}
 \caption{\label{fig7}
 Average coupling constant of spherical zinc nanoparticles without Coulomb repulsion. We also plot the extrapolation from the experimental bulk value, assuming $\mu^{*}=0.09$.\ \cite{anderson2}}
\end{figure}

\begin{figure}
 \includegraphics[width=1.0\textwidth]{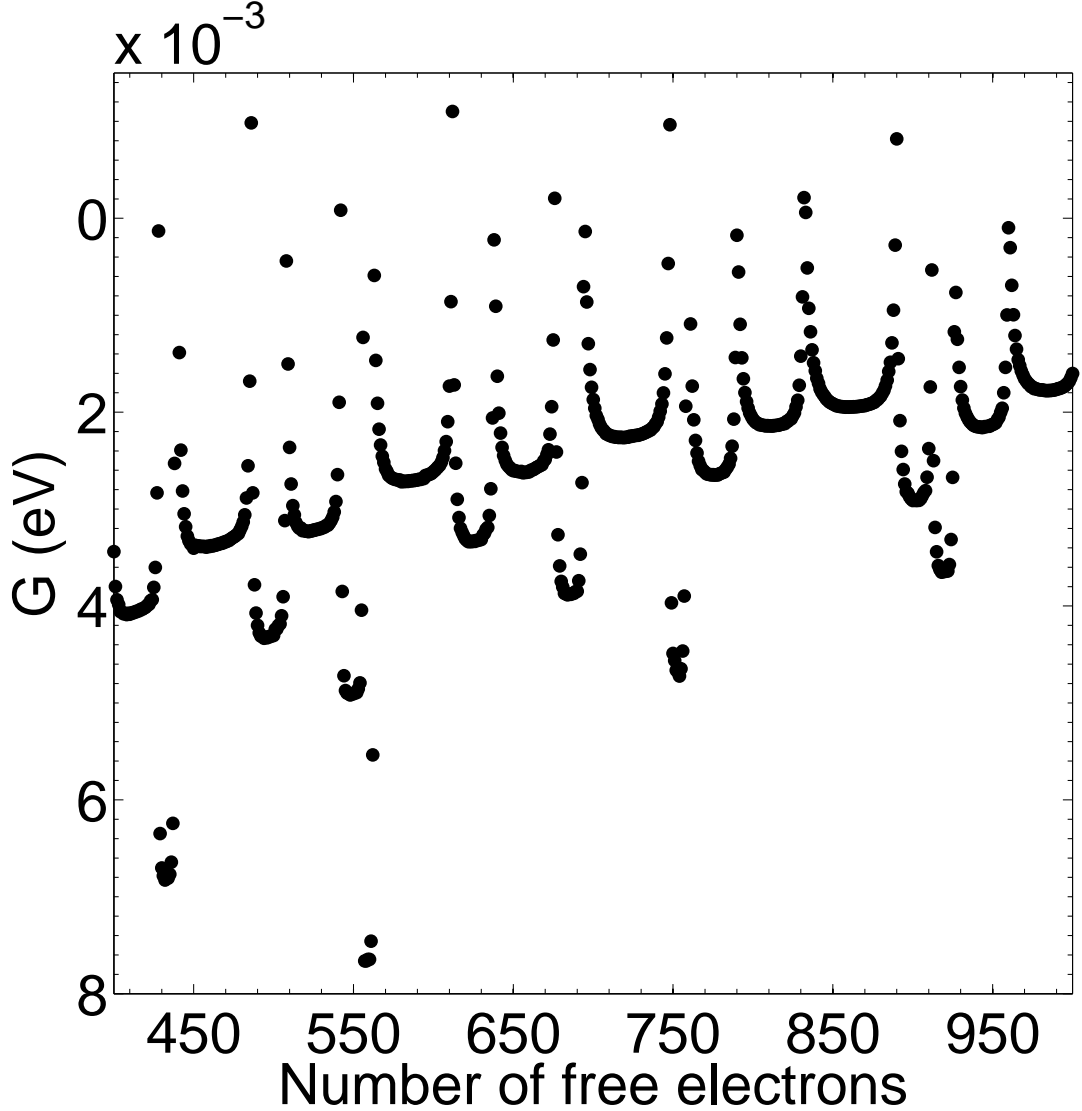}
 \caption{\label{fig8}
 Average coupling constant of deformed zinc nanoparticles with Coulomb interaction taken into account.}
\end{figure}

Unlike the potassium particles, the average phonon-mediated interaction in the zinc spherical particles is larger than the average Coulomb interaction. However, when both deformations and Coulomb interaction are taken into account we obtain repulsive interaction for almost all (99\%) zinc particles. The phonon-mediated average interaction of the spherical zinc particles is plotted in Fig.\ \ref{fig7} and the overall average interaction of the deformed zinc particles together with Coulomb interaction is shown in Fig.\ \ref{fig8}.

\subsection{The energy gap}\label{subsec:The energy gap}

\subsubsection{Energy gap of spherical particles}\label{subsubsec:Seniority gap}

\begin{figure}
 \includegraphics[width=1.0\textwidth]{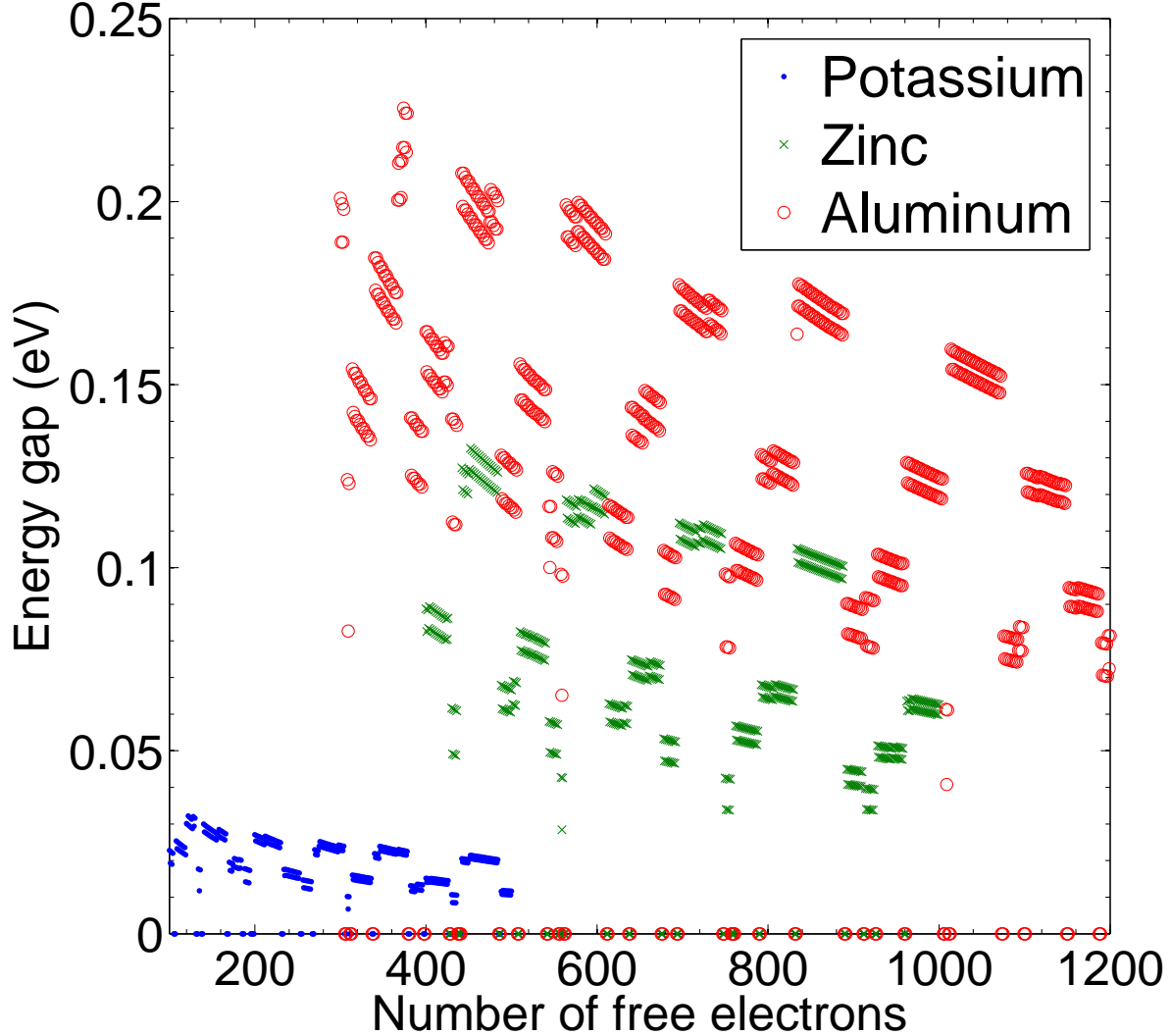}
 \caption{\label{fig9}
 (color online) Energy gap of spherical aluminum, zinc, and potassium nanoparticles plotted as a function of the number of free electrons in the nanoparticle, without Coulomb interaction, as obtained by either the seniority model or BCS model. The difference (equal to $G$) between the gap in particles with an even number of electrons in the HOS and particles with an odd number of electrons in the HOS is clearly seen for aluminum and zinc but is hard to observe for the potassium nanoparticles due to the relatively small average coupling constant. The larger jumps in the gap correspond to a transition from one shell to the next one. The transition between shells is marked by three particles with a zero gap, which correspond to $N_{HOS}=1,4l+1,4l+2$. The smaller jumps, such as the ones observed for the aluminum particles between $N_{e}=366$ and $N_{e}=367$ or $N_{e}=373$ and $N_{e}=374$, or for the zinc particles between $N_{e}=450$ and $N_{e}=451$ or $N_{e}=576$ and $N_{e}=577$, correspond to the addition of a phonon to the effective interaction within a given shell. Fig.\ \ref{fig10} shows a magnification of the region $N=400-500$ for aluminum.}
\end{figure}

\begin{figure}
 \includegraphics[width=1.0\textwidth]{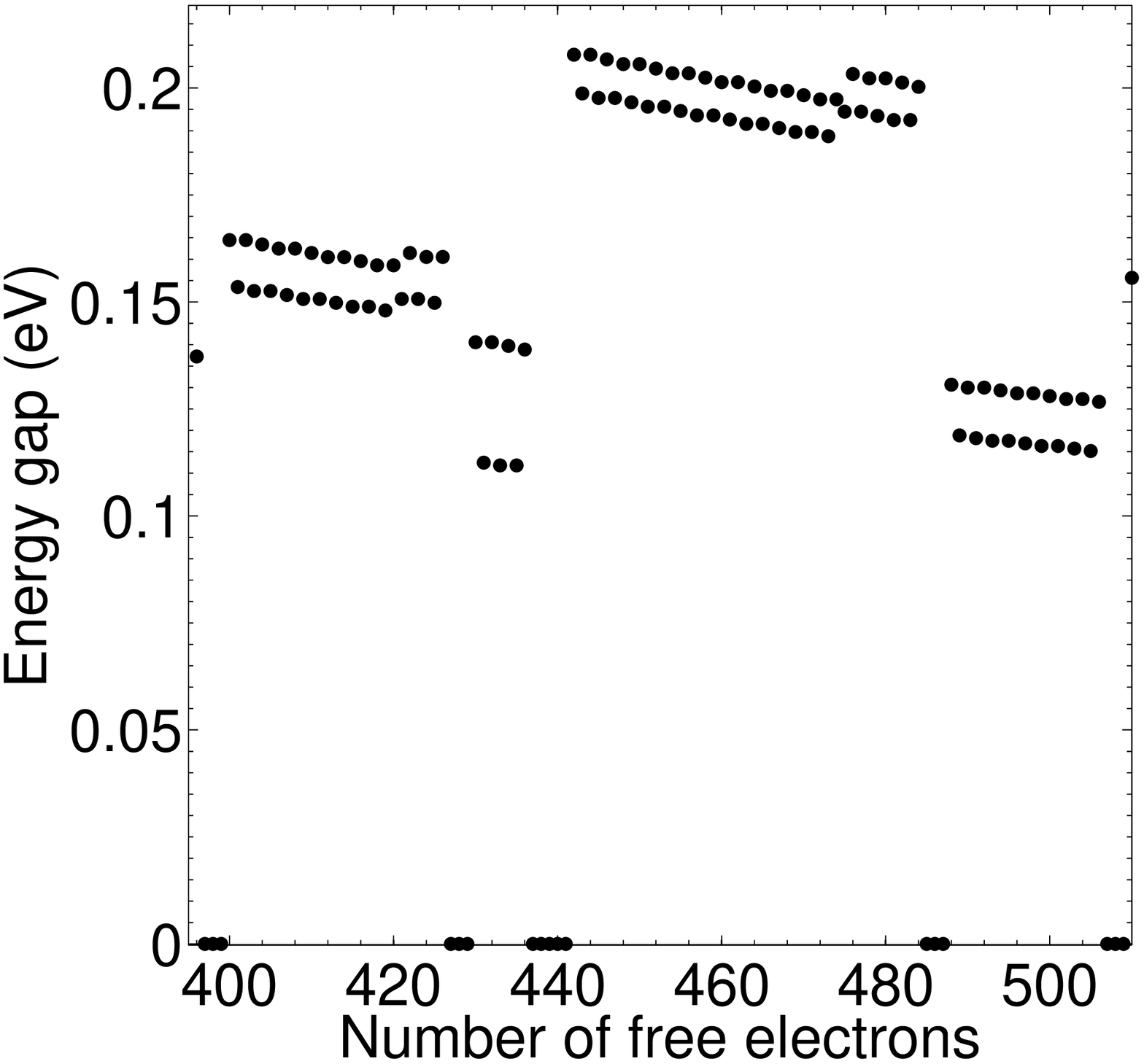}
 \caption{\label{fig10}
 Seniority model energy gap of spherical aluminum nanoparticles plotted as a function of the number of free electrons in the nanoparticle, without Coulomb interaction, as obtained by either the seniority model or BCS model. Five shells are shown -- $(l=7,n=1), (l=2,n=4),(l=0,n=5),(l=11,n=1),$ and $(l=5,n=3)$. The transition between shells is marked by three consecutive $N$ values for which the energy gap vanishes (within the framework of our model in which inter-shell interaction is ignored): $N_{HOS}=4l+1,4l+2$ for an almost-full or full shell and $N_{HOS}=1$ for the next shell. Around $N_{e}=440$ one notes 5 consecutive $N_{e}$ values with a zero energy gap. Three of these correspond to $N_{HOS}=4\times2+1,4\times2+2,1$ while the other two correspond to the shell $l=0,n=5$. The smaller jumps observed between $N_{e}=420$ and $N_{e}=421$ and between $N_{e}=473$ and $N_{e}=474$ are due to the addition of a single phonon to the effective interaction between the electrons and are not related to electronic shell effects.}
\end{figure}

In Fig.\ \ref{fig9} we plot the energy gap for spherical aluminum, zinc, and potassium nanoparticles, as calculated by the seniority model, without Coulomb interaction. The energy gaps are clustered according to the filling of the HOS of the nanoparticles. The gap shows an even-odd effect as a function of the number of electrons $N_{HOS}$ in the HOS: $\Delta=G(2l+1)$ or $\Delta=2Gl$ for even or odd $N_{HOS}$, respectively. As mentioned in section \ref{subsec:The BCS model}, the gap obtained by the BCS approximation coincides with the seniority gap for even $N_{HOS}$, regardless whether an average interaction or detailed matrix elements are used in the BCS calculation. The jumps in $G$, due to the addition of a single phonon to the effective interaction between the electrons, are magnified, by the multiplication of $G$ with half the HOS degeneracy. See, for example, in Fig.\ \ref{fig9} there is a jump in the gap between $N_{e}=366$ and $N_{e}=367$ or $N_{e}=373$ and $N_{e}=374$ for the aluminum particles; and between $N_{e}=450$ and $N_{e}=451$ or $N_{e}=576$ and $N_{e}=577$ for the zinc particles. Figure \ref{fig10} provides a magnified view of the region $N=400-500$ for aluminum, in order to demonstrate more clearly the jumps in the energy gap that are due to transitions into the next electronic shell and that are caused by the addition of a phonon to the effective interaction within a given shell.

The average energy gap of spherical aluminum particles is given in table \ref{average gaps}. In calculating the average gap for spherical particles we disregard particles with $N_{HOS}= 1, 4l+1,4l+2$. These particles may form a paired state only through the much weaker inter-shell interaction which we ignore. Therefore, they possess a smaller or even zero energy gap.

\subsubsection{Energy gap of deformed aluminum particles without Coulomb interaction}\label{subsubsec:The energy gap of deformed aluminum particles without Coulomb interaction}
The BCS model is used in order to estimate the energy gap for nanoparticles with non-degenerate HOS. Therefore, we calculate the energy gap only for particles with even $N_{HOS}$ and non-full HOS. The resulting average energy gaps of the aluminum particles for the various scenarios considered in this work are summarized in table \ref{average gaps}. In calculating the average energy gap of the deformed particles we take into account only particles with even $N_{HOS}$ and non-full HOS for which we were able to find a solution to the gap equations with negative condensation energy. The fraction of deformed particles (with and without Coulomb interaction) with even $N_{HOS}$ and non-full HOS for which we were able to find such a solution is shown in table \ref{average gaps}.

\begin{table}
 \begin{center}
 \begin{tabular}{|c|c|c|c|}\hline\hline
             & Average gap [eV]                       &  Fraction                          & $\left\langle\frac{gap}{LSEEE}\right\rangle$   \\ \hline\hline
 Spherical - 100\% bulk constants \hspace{1mm}    & $\hspace{5mm}0.139\hspace{5mm} $ & $\hspace{5mm}100\%\hspace{5mm} $ & $\hspace{5mm}-\hspace{5mm}$ \\
 Spherical - 75\% bulk constants \hspace{1mm}    & $\hspace{5mm}0.214\hspace{5mm} $ & $\hspace{5mm}100\%\hspace{5mm} $ & $\hspace{5mm}-\hspace{5mm}$ \\
 Deformation - 100\% bulk constants \hspace{1mm}   & $\hspace{5mm}0.040\hspace{5mm} $ & $\hspace{5mm}62\%\hspace{5mm} $ & $\hspace{5mm}7.0\hspace{5mm}$ \\
 Deformation+Coulomb - 100\% bulk constants \hspace{1mm}  & $\hspace{5mm}0.024\hspace{5mm} $ & $\hspace{5mm}51\%\hspace{5mm} $ & $\hspace{5mm}4.1\hspace{5mm}$ \\
 Deformation+Coulomb - 75\% bulk constants \hspace{1mm}  & $\hspace{5mm}0.045\hspace{5mm} $ & $\hspace{5mm}50\%\hspace{5mm} $ & $\hspace{5mm}6.6\hspace{5mm}$ \\
 \hline\hline
 \end{tabular}
 \end{center}
 \begin{center}
 \begin{minipage}{0.7\textwidth}
 \caption{\label{average gaps}
 Summary of the results we obtained for the energy gap of the aluminum particles in the various scenarios considered in this work. In the second column, we show the average energy gap of the aluminum particles. In the third column, we present the fraction of the aluminum particles for which we were able to find a solution to the gap equation with negative condensation energy. In the fourth column, the average ratio between the energy gap of the deformed aluminum particles and the lowest single electron excitation energy (LSEEE) is shown.}
 \end{minipage}
 \end{center}
\end{table}

The energy gap is seven times larger on average than the LSEEE of the deformed aluminum particles without Coulomb interaction. In calculating this average ratio (as well as the same ratio when Coulomb interaction is taken into account) we take into account only particles with even $N_{HOS}$ and non-full shells for which we find a negative condensation energy. Of those, we disregard the particles for which the LSEEE involves a shell transition.

\begin{figure}
 \includegraphics[width=1.0\textwidth]{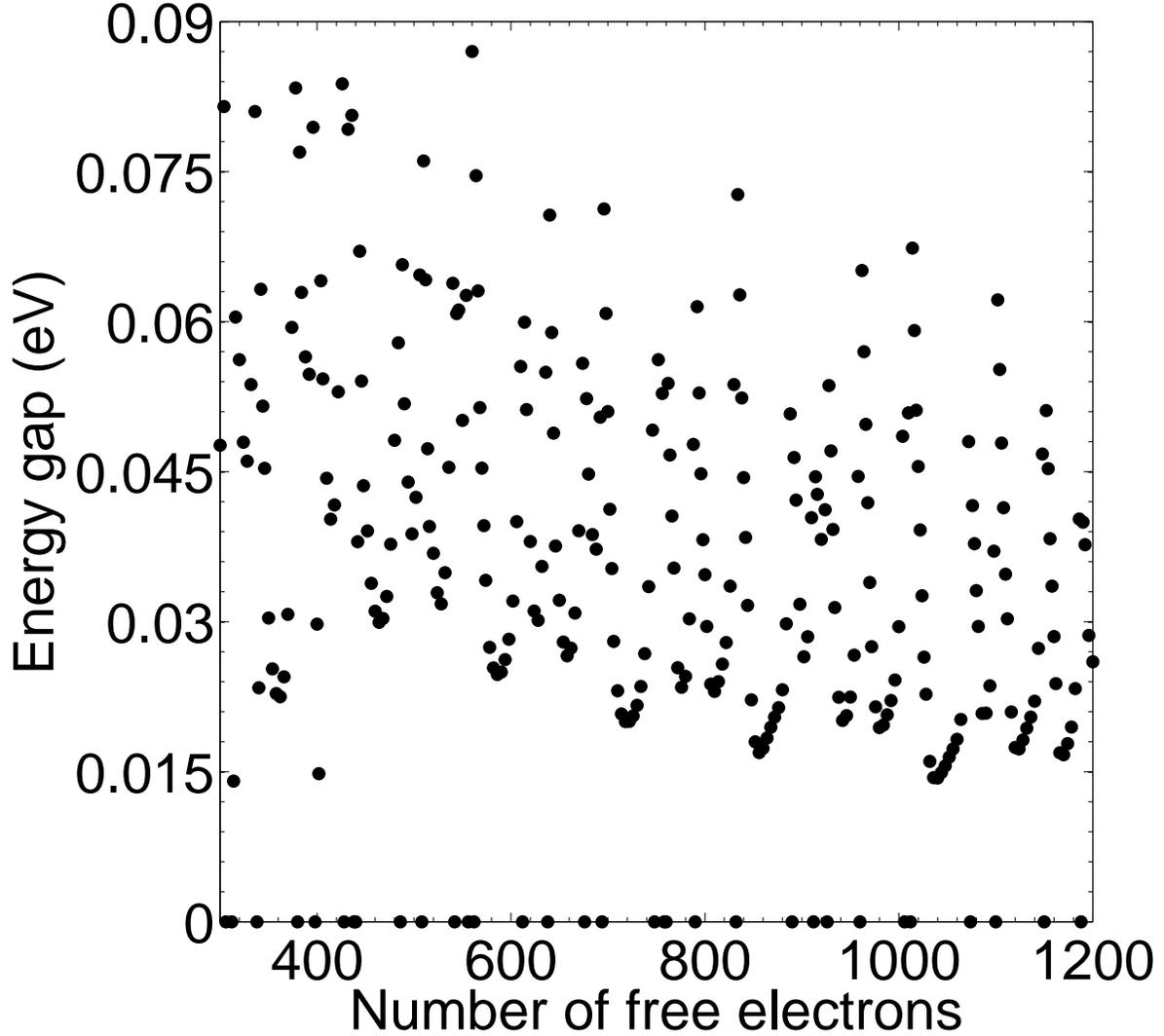}
 \caption{\label{fig11}
 BCS model gap of deformed aluminum nanoparticles, without Coulomb interaction, with even $N_{HOS}$ and negative condensation energy plotted as a function of the number of free electrons in the particle. The gap is calculated using the interaction matrix elements $G_{mm^{'}}$, and by solving Eqs.\ \eqref{eq:gap T0} and \eqref{eq:number of electrons}.}
\end{figure}

\begin{figure}
 \includegraphics[width=1.0\textwidth]{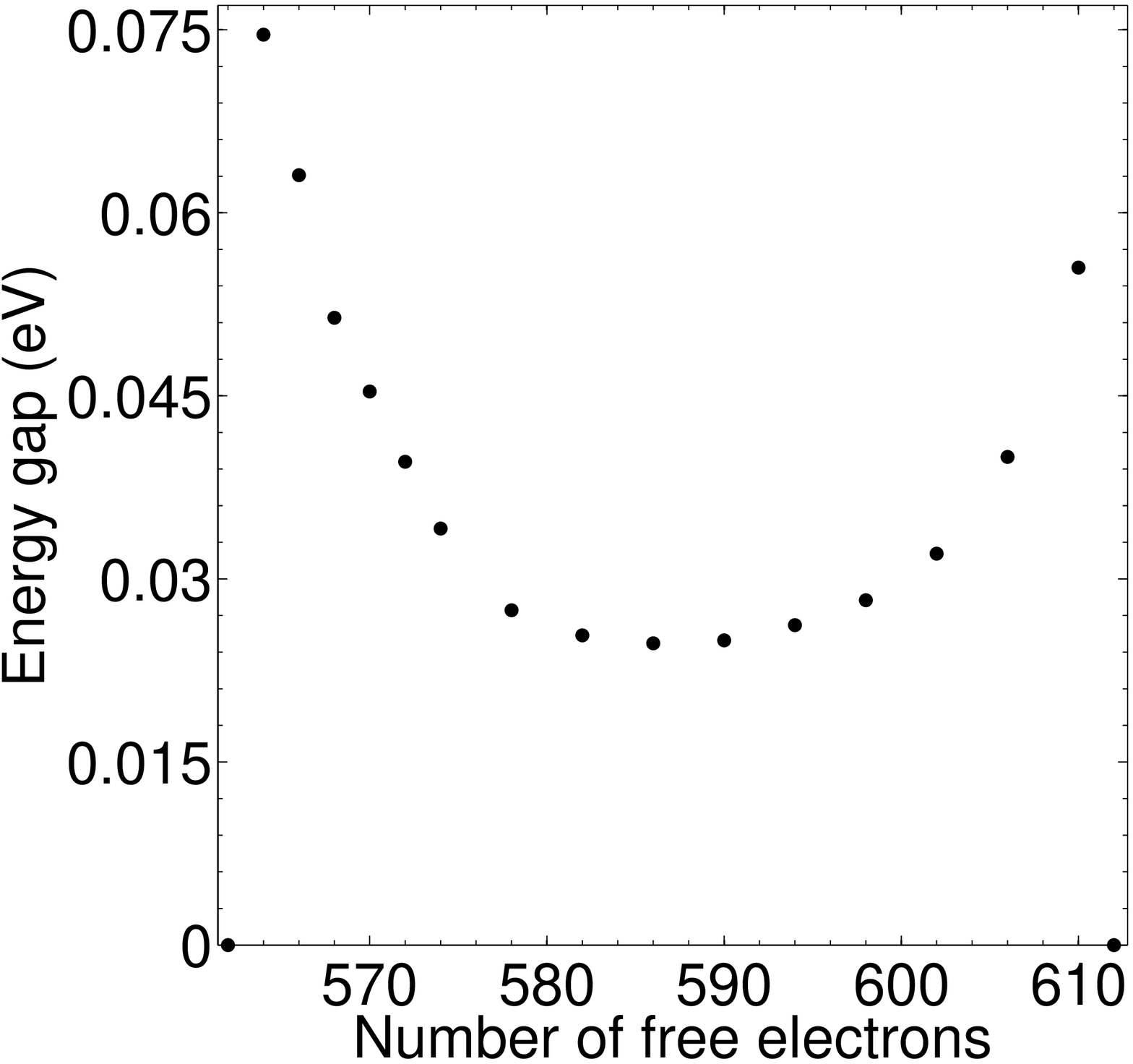}
 \caption{\label{fig12}
 BCS model gap of deformed aluminum nanoparticles, without Coulomb interaction, with even $N_{HOS}$ varying between $N_{HOS}=562$ to $N_{HOS}=612$ and negative condensation energy. All particles belong to the HOS $l=12, n=1$. The gap is calculated using the interaction matrix elements $G_{mm^{'}}$, and by solving Eqs.\ \eqref{eq:gap T0} and \eqref{eq:number of electrons}.}
\end{figure}

In Figs.\ \ref{fig11} and \ref{fig12}, we plot the resulting gap for deformed aluminum nanoparticles without Coulomb interaction. We plot the energy gap that corresponds to a solution of the gap equations with the lowest negative condensation energy. We note that, for a large portion of the deformed particles, a part of the gap parameters $\Delta_{m}$ is essentially equal to zero (within the accuracy of our numerical solution). However, this is an artifact that disappears with the addition of Coulomb interaction, which results in gap parameters that are both positive and negative.

As can be seen from table \ref{average gaps} we find solutions to the gap equations with negative condensation energies for most nanoparticles with non-degenerate shells. Also, as can be seen from Fig.\ \ref{fig11} and more clearly from Fig.\ \ref{fig12}, the energy gap generally exhibits a quite smooth behavior as a given energy shell is been filled up, and a minimum near half-shell fillings. This behavior is consistent with the behavior of the average coupling constants (Fig.\ \ref{fig1}). However, we fail to obtain solutions to the gap equations with negative condensation energy for some particles, which many times are located near shell opening or shell closing. These failures are probably due to the limitations of our numerical procedure and not due to a real feature of our system. We claim that these are probably failures of our numerical procedure and not a real feature of our system because for most energy shells we are able to find appropriate solutions for particles located near half-shell filling, where the shell splitting is maximal and the formation of a paired state is least likely. We note that although the general structure shown in Figs.\ \ref{fig11} and \ref{fig12} seems to be reasonable, we cannot prove that solutions with lower condensation energies do not exist.

\subsubsection{Energy gap of deformed aluminum particles with Coulomb interaction}\label{subsubsec:The energy gap of deformed aluminum particles with Coulomb interaction}
\begin{figure}
 \includegraphics[width=1.0\textwidth]{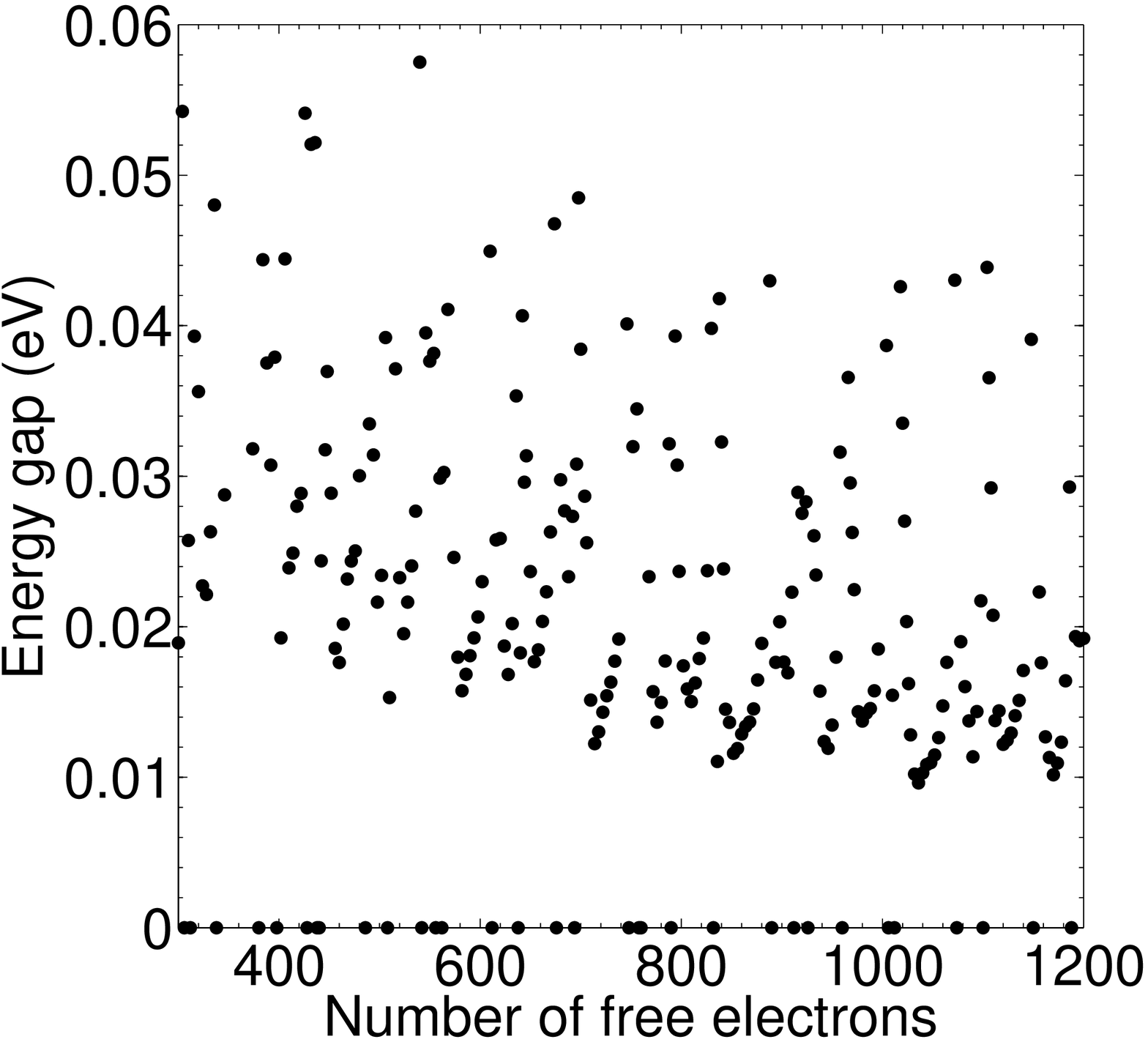}
 \caption{\label{fig13}
 BCS model gap of aluminum nanoparticles when both deformations and Coulomb repulsion are taken into account plotted as a function of the number of free electrons in the particle. Only particles with even $N_{HOS}$ and negative condensation energy are shown. The gap is calculated using the detailed interaction matrix elements $G_{mm^{'}}$, and by solving Eqs.\ \eqref{eq:gap T0} and \eqref{eq:number of electrons}.}
\end{figure}

\begin{figure}
 \includegraphics[width=1.0\textwidth]{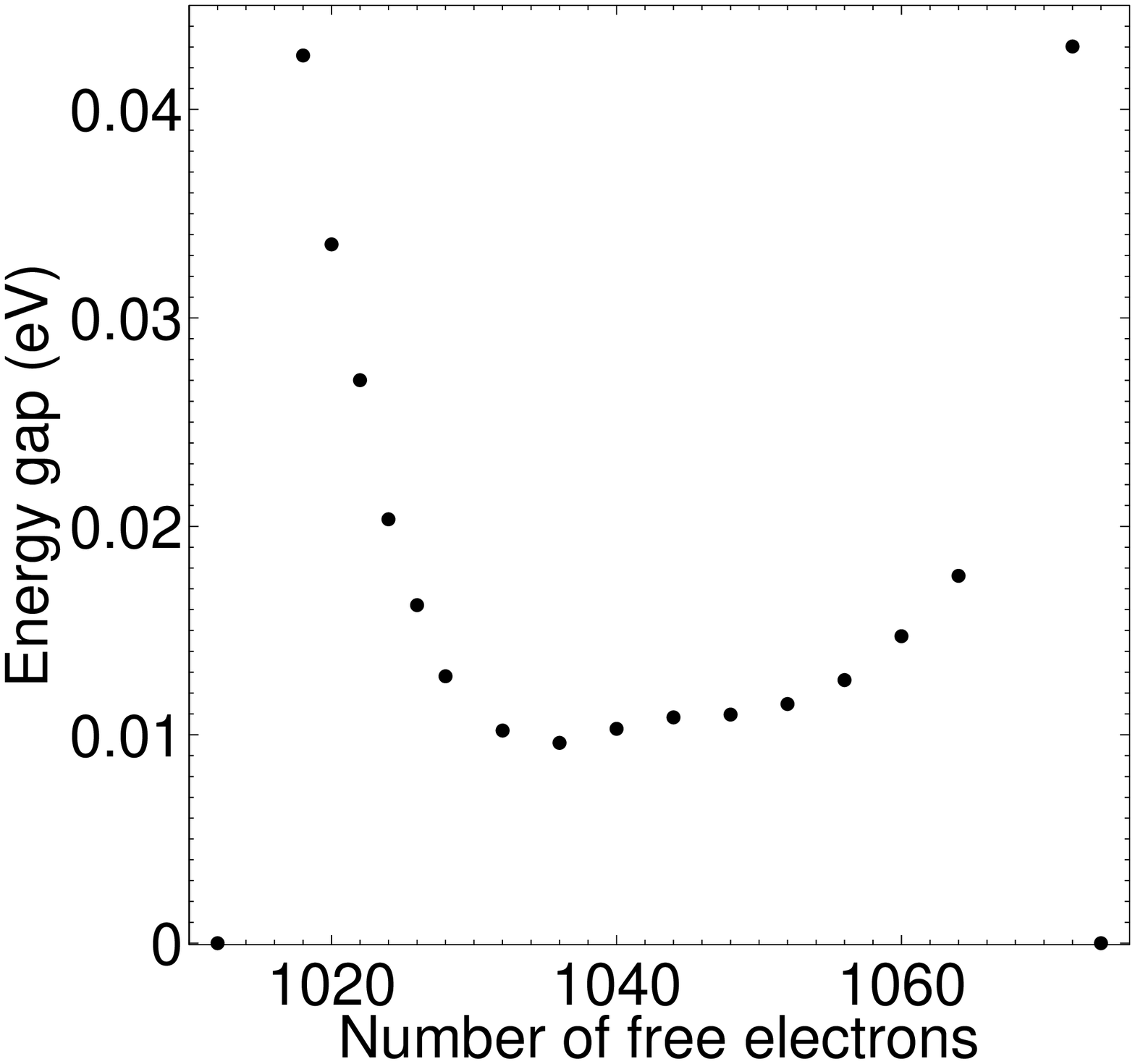}
 \caption{\label{fig14}
 BCS model gap of aluminum nanoparticles when both deformations and Coulomb repulsion are taken into account plotted as a function of the number of free electrons in the particle. Only particles with negative condensation and even $N_{HOS}$ varying between $N_{HOS}=1012$ to $N_{HOS}=1074$ are shown. All particles belong to the HOS $l=15, n=1$. The gap is calculated using the detailed interaction matrix elements $G_{mm^{'}}$, and by solving Eqs.\ \eqref{eq:gap T0} and \eqref{eq:number of electrons}.}
\end{figure}

In Figs.\ \ref{fig13} and \ref{fig14}, we show the energy gap for aluminum particles when Coulomb interaction is added on top of deviations from spherical symmetry. The average value of the energy gap is given in table \ref{average gaps}. We find a solution to the gap equations with negative condensation energy for 50\% of the aluminum particles with even $N_{HOS}$. For these particles, we find that the energy gap is approximately 4 times larger on average than the LSEEE. However, it should be noted that near shell opening or closing, where the effect of deformations is minimal, the gap can be larger than the LSEEE by a factor of 10 or more. Also, the average energy gap is still two orders of magnitude larger than the energy gap found in superconducting bulk aluminum \cite{delft1} (which is about 0.18meV), but smaller (by an order of magnitude) than the mean level spacing between the HOS and the LUS (see table \ref{properties} for the average level spacing between the HOS and the LUS in the aluminum particles ).

Aluminum particles with smaller Lam\'{e} coefficients exhibit a larger energy gap on average than the aluminum particles with bulk  Lam\'{e} coefficients (see table \ref{average gaps}). The percentage of particles for which we are able to find solutions to the gap equations with negative condensation energy is similar to the one obtained for particles with bulk Lam\'{e} coefficients.

\subsubsection{Gap parameters anisotropy in aluminum particles}\label{subsubsec:Gap parameters anisotropy}

The calculated values of the gap parameters $\Delta_{m}$ exhibit a large anisotropy compared to bulk aluminum.  The average ratio between the maximal and minimal absolute values of $\Delta_{m}$ is about 10 when both deformations and Coulomb interaction are taken into account. By comparison, according to the calculations of Leavens and Carbotte \cite{leavens}, the corresponding ratio in bulk superconducting aluminum is about 1.4. We note that Croitoru et al.\ \cite{croitoru} have found a large spatial anisotropy in the gap parameter of spherical nanoparticles, calculated by solving the Bogoliubov-de Gennes equations.

\subsubsection{Energy gap of zinc and potassium particles}\label{subsubsec:Energy gap of the zinc and potassium particles}
\begin{figure}
 \includegraphics[width=1.0\textwidth]{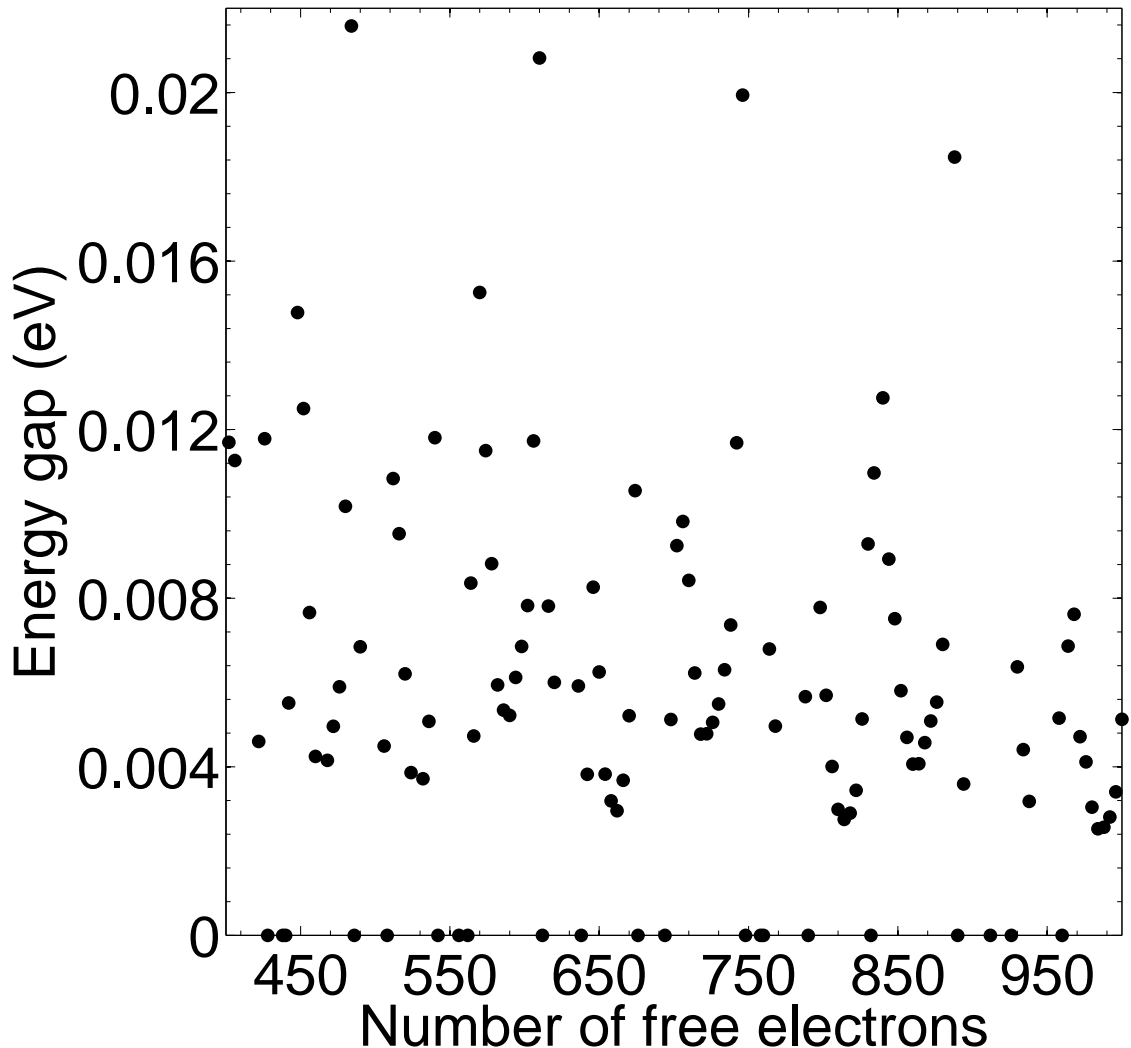}
 \caption{\label{fig15}
 BCS model gap of zinc nanoparticles, when both deformations and Coulomb repulsion are taken into account, plotted as a function of the number of free electrons in the particle. Only particles with even $N_{HOS}$ and negative condensation energy are shown. The gap is calculated using the detailed interaction matrix elements $G_{mm^{'}}$, and by solving Eqs.\ \eqref{eq:gap T0} and \eqref{eq:number of electrons}.}
\end{figure}

In Fig.\ \ref{fig15} we show the energy gap of zinc particles when deformations are taken into account together with Coulomb repulsion. Although zinc particles are less susceptible to deformations than aluminum particles, the smaller electron-phonon interaction and the resulting effective electron-electron interaction lead to a lower average energy gap for zinc particles than for aluminum particles. We find that the average energy gap in  spherical zinc particles without Coulomb interaction is equal to 0.08eV, and to 0.007eV when both deformations and Coulomb interaction are taken into account. We find a solution to the gap equation with negative condensation energy only in 37\% of the zinc particles (with even $N_{HOS}$ and non-full HOS). We are unable to find solutions to the gap equations for potassium nanoparticles when both deformations and Coulomb interaction are taken into account. This implies that within the framework of the BCS grand-canonical approximation, pairing correlations are completely destroyed in potassium nanoparticles.

\section{Possible measurable quantities}\label{sec:Additional possible measurable quantities}

In bulk superconductors, the energy gap is directly observable by measuring the electronic DOS of the material (for example, by tunneling experiments), which is significantly altered compared to the normal state. By contrast, there is no qualitative difference between the spectrum of unpaired electrons in ultra-small nanoparticles and the spectrum of paired electrons, because the spectrum is discrete in both cases. As mentioned above, we anticipate that the lowest excitation energy of the paired state (\textit{i.e.}\ the energy gap) to be several times larger on average than the lowest excitation energy of the unpaired state, and smaller, on average, by an order of magnitude than the energy difference between the HOS and the LUS. Thus, an observation of excitation energies that lie between the HOS-LUS difference and the energy differences expected from the unpaired shell structure may indicate the presence of pairing in ultra-small nanoparticles.  The effect of the HOS splitting is minimal near shell opening or closing, while our results are more accurate near shell closing since the number of particles participating in the paired state is maximal. Therefore, like Kresin and Ovchinnikov \cite{kresin3} we suggest searching for evidence for pairing correlations in nanoparticles with $N_{HOS}$ near shell closing.

Although observation of alteration in the electronic spectrum may be used as an indicator of the presence of pairing, one must remember that our description of the single-electronic shell structure is quite crude and neglects effects that may contribute to the splitting of the HOS, such as non-axially symmetric deformations, surface roughness and disorder. Therefore, it is possible that we underestimate the magnitude of the splitting, and as a result overestimate the magnitude of the energy gap. If this is indeed the case it will be hard to distinguish between splitting caused by pairing and splitting caused by deviations from the spherical shell structure.

The seniority model predicts [Eq.\ \eqref{eq:delta E S to S plut two}] that the spacing between adjacent energy levels decreases linearly when higher levels within the paired HOS are considered. The presence of such a structure in the electronic energy spectrum could indicate the existence of pairing interaction. Also, according to the seniority model the difference between two adjacent level spacings is equal to $2G$ (\textit{i.e.}\ to a few meV). These two predictions are somewhat modified (but not substantially) when the effects of deformations are taken into account within the framework of the BCS approximation.

An alternative fingerprint of the presence of pairing could be found in magnetization measurements. Let us assume that a nanoparticle does not exhibit pairing correlations and that the degeneracy of the HOS is completely lifted (except for the spin degeneracy) due to deformations. This is the accepted description of the energy levels of atomic clusters.\ \cite{heer1,heer2} We further assume that the energy splitting within a sub-shell (\textit{i.e.}\ between $m$ and $-m$) is smaller than the splitting between sub-shells (\textit{i.e.}\ between different values of $|m|$), and therefore the electrons first fill up the sub-shells with small values of $|m|$. We focus on nanoparticles with two electrons in the highest populated sub-shell (\textit{i.e.}\ half-filling of the sub-shell) and consider the zero-temperature value of the particle magnetization. The magnetic field inducing the magnetization is taken to be constant, directed in the $z$ direction, and weak enough so it does not change the electron population in the various sub-shells. The effects of deformations or renormalization on the electronic wave functions are ignored.

Under the above assumptions we find, to lowest order in the weak magnetic field, that the magnetization of the nanoparticle is
\begin{equation}\label{eq:magnetization}
M=-\mu_{B}m,
\end{equation}
where $\mu_{B}$ is the Bohr magneton, and we assume that the lowest non-full energy sub-shell corresponds to $-m$. By contrast, if pairing is present in the same nanoparticle, then the magnetization will be zero. However, it should be noted that if the number of electrons in the particle corresponds to a full highest occupied sub-shell, then the magnetization is zero (to the first order in the field strength) also in the absence of pairing.

A different approach may rely on the detection of the transition between the paired and the unpaired state as the temperature is varied. Cao et al.\ \cite{cao} measured the heat capacity of aluminum atomic clusters containing between 43 and 48 atoms. They found a peak in the heat capacity of clusters containing 44 and 47 atoms around $T=200K$. These peaks are claimed to represent the transition between the paired and the unpaired state as the temperature of the clusters is varied. The experimental results are somewhat higher, but are still in accordance with the calculations of Kresin and Ovchinnikov.\ \cite{kresin1,kresin2,kresin3} We note that when evaluating the energy gap of the nanoparticles at such a temperature, it may be necessary to take into account finite temperature broadening of the sharp, almost degenerate, single-electron levels.

\section{Summary}\label{sec:Summary}
We investigated the possibility of pairing interaction in metallic nanoparticles containing a few hundreds of atoms at zero-temperature. Three materials were considered -- aluminum, zinc and potassium. We started from a simple model of non-interacting phonons (quantized normal modes of vibration of a stress-free elastic sphere) and electrons (fermions in a spherical potential box) in the nanoparticles. We introduced an electrostatic screened Thomas-Fermi interaction potential between the electrons and the phonons, using the Thomas-Fermi screening length in bulk materials. We then derived an effective phonon-mediated electron-electron interaction, resulting from the underlying electron-phonon interaction. We found that the strongest attractive interaction was between time-reversed electron pairs within the same energy shell. We neglected the rest of the effective interaction and assumed a model pairing Hamiltonian consisting only of the interaction between electrons in the HOS.

The effective interaction was derived by means of either the Fr\"{o}hlich transformation \cite{frohlich1} or the similarity renormalization. \cite{glazek,mielke1,mielke2} The application of the Fr\"{o}hlich transformation was more straightforward but restricted to particles with degenerate HOS. However, the degeneracy of the HOS is lifted due to deviations from spherical symmetry. The effects of deformations on the effective interaction between the electrons were taken into account within the framework of the similarity renormalization method. Our application of this technique followed the work of Mielke.\ \cite{mielke1} However, the interaction we obtained includes a different cutoff function than the one obtained by Mielke.\ \cite{mielke1} Our cutoff function, which resembles more the result obtained by Hubsch and Becker \cite{hubsch}, reflects the fact that a phonon cannot mediate the interaction between electron states with an energy separation larger than its own energy. We also included a screened Coulomb interaction in our model Hamiltonian, and found that it plays a significant role in reducing pairing effects.

Unlike the Fr\"{o}hlich interaction, the effective interaction obtained by the application of the similarity transformation depends on the renormalized electron energies due to the electron-phonon interaction. The size of the renormalization was also obtained using the similarity renormalization method. However, within the framework our model, in which inter-shell interactions are neglected, the renormalization is unimportant since (for spherical particles) it shifts all of the states within the HOS by the same amount. However, renormalization may play a more important role when inter-shell interactions are taken into account, especially when the non-renormalized HOS and LUS are nearly degenerate.

On average, there was no large difference between the average phonon-mediated interaction we obtained and the extrapolation from bulk interaction. However, finite size effects induced variation in the values of the average interaction in contrast to the smooth monotonic behavior of the extrapolation. Finite size effects also caused a large scatter between the detailed effective interaction matrix elements for a given particle. This scatter was further enhanced in deformed nanoparticles, and it was important to take it into account when we evaluated the energy gap in those particles. We also found that the entrance of an additional single phonon to the effective interaction between the electrons resulted in small jumps in the average effective interaction, and in more pronounced jumps in the energy gap of spherical particles. However, it was hard to observe the fingerprints of this effect in the calculated energy gap of deformed nanoparticles.

The effect of the pairing Hamiltonian on the electronic spectrum was evaluated by using the seniority model (when we considered particles with degenerate HOS) or by solving the BCS gap equation for the HOS. When Coulomb interaction and deformations were ignored, we found that aluminum particles exhibit an average energy gap (\textit{i.e.}\ the lowest excitation energy of the paired electrons) of about 0.14eV, while the average gap in zinc and potassium particles was found to be 0.07eV and 0.02eV respectively. The addition of Coulomb interaction, together with deformations, reduced the energy gap by a factor of 5.5 on average in the aluminum particles and by a factor of 10 in the zinc particles. On the other hand, the same effects reduced the energy gap to zero in the potassium particles. Despite the large reduction, our calculations indicate that a large portion of both aluminum and zinc particles should exhibit pair correlations and modifications in their electronic spectrum.

We found that within the framework of our approximate model, the resulting energy gap is on average intermediate between the LSEEE and the energy difference between the HOS and the LUS. Additionally, if pairing is present, then we expect that the magnetization of nanoparticles with certain fillings of the HOS (namely, two electrons at the highest sub-shell within the HOS) to be different from the magnetization of the unpaired ground-state. Therefore, magnetization may serve as an additional fingerprint of the existence of pairing correlations in metallic nanoparticles.

Our results indicate that the size (and maybe even the existence) of the modifications in the electronic energy spectrum are sensitive to several factors such as the effective elastic constants of the particles, the details of the energy splitting of the HOS (which may be affected by additional mechanisms such as disorder, non-axially symmetric deformations, and surface roughness), and maybe the detailed description of the Coulomb interaction. Therefore, it is probably necessary to consider these aspects more accurately in order to reliably predict whether pairing correlations do exist in specific metallic nanoparticles.

An additional improvement involves applying a canonical-ensemble treatment of the pairing Hamiltonian instead of the grand-canonical BCS approximation. The simplest fixed-N treatment for the energy gap \cite{braun2,braun3} is obtained by projecting the BCS ground state, as well as the BCS ground state with two ``blocked'' single electron states, onto the subspace of states with a fixed number of electrons in the HOS and calculating the energy difference between them. This method yields results that are quite close to the ones obtained by an exact solution of the pairing Hamiltonian for aluminum particles with constant interaction strength\cite{braun3,sierra} and uniform or random level structure. Therefore, it is probably sufficient in order to evaluate the accuracy of the BCS approximation results for our case, where an exact approach is inapplicable. More sophisticated methods could also be used, but their application should be considered in light of the approximations involved in deriving the effective interaction between the electrons and in estimating the single-electron level structure.

Finally, we note that the standard approach for calculating properties of superconducting bulk material relies either on the application of Eliashberg theory \cite{mcmillan,allen3,allen2,carbotte} or density functional theory \cite{oliveira,suvasini,kurth,luders,marques} instead of a BCS approach based on an effective model Hamiltonian. In fact, density functional theory was already used to describe pairing in the nucleus \cite{perlifmmode,dobaczewski} and in nanometric superconductors.\ \cite{lacroix} Specifically, the application of density functional approach to equally spaced \cite{lacroix,hupin} or randomly distributed \cite{hupin} electronic spectrum (both with constant pairing interaction) reproduced remarkably well the results of the exact Richardson solution. It would be interesting to see how our simple BCS approach would fair compared to a density functional approach applied to non-constant pairing interaction and an approximate electronic shell structure. On the other hand, the adaption of the Eliashberg theory to the problem of pairing in ultra-small nanoparticles will enable an alternative (and indeed a far more widely used one in the context of bulk material) to our similarity renormalization + BCS solution approach, for the calculation of pairing correlations from the  underlying electron-phonon interaction.

\appendix

\section{Spheroidal phonons and their longitudinal component}\label{app:The longitudinal phonon sub-system}

The details of the solution of the linear elasticity equation of motion can be found in many references \cite{eringen}, here we shall cite only the relevant results.

The frequencies of the spheroidal modes are determined by solving
\begin{equation}\label{eq:deteminant}
T_{11}T_{43}-T_{41}T_{13}=0,
\end{equation}
for $\l\neq 0$, where
\begin{eqnarray*}
 & &T_{11}=\left (l^{2}-l-\frac{\eta^{2}}{2}\right )j_{l}(\xi)+2pRj_{l+1}(\xi) \label{eq:t11}\ ,\\
 & &T_{13}= l(l+1)[(l-1)j_{l}(\eta)-\eta j_{l+1}(\eta)] \label{eq:t13}\ ,\\
 & &T_{41}=(l-1)j_{l}(\xi)-\xi j_{l+1}(\xi)\label{eq:t41} \label{eq:t41}\, \\
 & &T_{43}=\left (l^{2}-\frac{\eta^{2}}{2}-1\right )j_{l}(\eta)+\eta
 j_{l+1}(\eta) \label{eq:t43}\,
\end{eqnarray*}
where $j_{l}$ is the spherical Bessel function of the order $l$.
For $l=0$ one needs to solve
\begin{equation}\label{eq:detrminant l=0}
T_{11}=0.
\end{equation}
In the above equations $\eta=pR$ and $\xi=\frac{\omega}{c_{t}}R$, where $c_{t}$ is the transverse sound velocity.

The ratio between the amplitude of the longitudinal component $A$ and the amplitude of the transverse component $C$ is determined by the stress-free boundary conditions imposed on the surface of the sphere ($\sigma_{rr}=\sigma_{\theta r}=\sigma_{\phi r}=0$, where $\sigma_{ij}$ are the components of the stress-tensor). If $l>0$ the ratio is given by $-T_{11}/T_{13}$, while if $l=0$ then $C=0$ and $A$ is determined solely by the wave function normalization condition.

\section{Derivation of the electron-phonon matrix elements}\label{app:The detailed derivation}

We derive the explicit expression for the electron-phonon matrix elements $M_{LMN}$ \eqref{eq:defnition of M}. Using Eqs.\ \eqref{eq:electrons wave}, and \eqref{eq:u longitudinal}-- \eqref{eq:quantization vibration}, we obtain the following expression for the matrix elements
\begin{eqnarray}
M_{LMN}=\int\int & &
B_{l_{1}n_{1}}j_{l_{1}}(k_{l_{1}n_{1}}r_{1})Y^{*}_{l_{1}m_{1}}(\Omega_{1})B_{l_{1}n_{1}}j_{l_{2}}(k_{l_{2}n_{2}}r_{1})Y_{l_{2}m_{2}}(\Omega_{1})
\frac{e^{-k_{TF}\left |\mathbf{r_{1}}-\mathbf{r_{2}}\right
|}}{\left|\mathbf{r_{1}}-\mathbf{r_{2}}\right|}\nonumber\\
& &A_{l_{3}n_{3}}\sqrt{\frac{\hbar\omega_{l_{3}n_{3}}}{2\rho c_{lo}^{2}}} j_{l_{3}}(k_{l_{3}n_{3}}r_{2})Y_{l_{3}m_{3}}(\Omega_{2})
r_{1}^{2}dr_{1}d\Omega_{1}r_{2}^{2}dr_{2}d\Omega_{2},\label{eq:hamiltonian1}
\end{eqnarray}
where the indexes $LMN$ denote the same as in \eqref{eq:spesicifc e-p 2}. We insert the spherical harmonic expansion of the
screened electrostatic potential \eqref{eq:screened potential}, and the explicit expression for $B_{ln}$ \eqref{eq:electrons normalization} into
\eqref{eq:hamiltonian1}, and obtain
\begin{eqnarray}
M_{LMN}=\sum_{l=0}^{\infty}\sum_{m=-l}^{l}& &\frac{8\pi ze^{2}n_{0}k_{TF}}{R^{3}}\sqrt{\frac{\hbar\omega_{l_{3}n_{3}}}{2\rho
c_{lo}^{2}}}\frac{A_{l_{3}n_{3}}}{j_{l_{1}+1} \left(a_{l_{1}n_{1}}\right)j_{l_{2}+1}\left(a_{l_{2}n_{2}}\right)}\nonumber \\
& &\left\{\int^{R}_{0}dr_{1}r_{1}^{2}j_{l_{1}}\left(k_{1}r_{1}\right)j_{l_{2}}\left(k_{2}r_{1}\right)\left[k_{l_{3}}\left(k_{TF}r_{1}\right)
\int^{r_{1}}_{0}dr_{2}r_{2}^{2}j_{l_{3}}\left(k_{3}r_{2}\right)i_{l_{3}}\left(k_{TF}r_{2}\right)\right.\right.\nonumber \\
&&\left.\left.+i_{l_{3}}\left(k_{TF}r_{1}\right)\int^{R}_{r_{1}}dr_{2}r_{2}^{2}j_{l_{3}}\left(k_{3}r_{2}\right)k_{l_{3}}
\left(k_{TF}r_{2}\right)\right]\right\}\nonumber\\
& &\int Y^{*}_{l_{1}m_{1}}(\Omega_{1})Y_{l_{2}m_{2}}(\Omega_{1})Y_{lm}(\Omega_{1})d\Omega_{1} \int
Y_{l_{3}m_{3}}(\Omega_{2})Y_{lm}(\Omega_{2})d\Omega_{2}\label{eq:hamiltonian2}\ .
\end{eqnarray}
The angular integrations in \eqref{eq:hamiltonian2} yield
\begin{eqnarray}
& &\int Y_{l_{3}m_{3}}(\Omega_{2})Y_{lm}(\Omega_{2})d\Omega_{2}=\delta_{mm_{3}}\delta_{ll_{3}}\label{eq:conservation of m}\\
& &\int Y^{*}_{l_{1}m_{1}}(\Omega_{1})Y_{l_{2}m_{2}}(\Omega_{1})Y_{lm}(\Omega_{1})d\Omega_{1}=\nonumber \\
& &\sqrt{\frac{\left(2l_{1}+1\right)\left(2l_{2}+1\right)}{4\pi\left(2l+1\right)}}
c\left(l_{1},l_{2},l;0,0,0\right)c\left(l_{1},l_{2},l;m_{1},-m_{2},m\right)(-1)^{m_{2}}\label{eq:conservation of l}\ .
\end{eqnarray}
We note that the only non-vanishing Clebch--Gordan coefficients in \eqref{eq:conservation of l} are those complying with the conservation laws
(\ref{eq: usual restriction on l}--\ref{eq:conservation of mz}). Therefore the summation in (\ref{eq:hamiltonian2}) is restricted only to these
values of $l$'s, and $m$'s.

\section{Application of the similarity renormalization}\label{app:detailed}

The similarity renormalization is a renormalization scheme in which an initial non-diagonal Hamiltonian, which connects states separated by a large energy difference (up to a large initial cutoff $\Lambda$), is transformed, via a continuous set of infinitesimal unitary transformations, into a band diagonal effective Hamiltonian with a smaller energy cutoff $\lambda$. The renormalization is carried out perturbatively and in a manner which avoids small-energy denominators that may appear in ordinary perturbation expansions, or in a single unitary transformation of the initial Hamiltonian such as the Fr\"{o}hlich transformation (for more details see \cite{glazek,mielke1}).

Guided by the results obtained using the Fr\"{o}hlich transformation, and in order to simplify the notation, we concentrate on the interaction within the HOS and ignore inter-shell interaction. The entire Hamiltonian can be obtained following the same steps we outline below (see also the results of Mielke \cite{mielke1} who derived a general effective electron-electron interaction for the bulk system).

The transformed Hamiltonian $H_{\lambda}$ can be divided into a diagonal part $H_{0\lambda}$ and an interaction part $H_{I\lambda}$. It is preferable to carry out the pertubative renormalization expansion using only normal ordered operators.\ \cite{kehrein} Therefore, we write the $H_{0\lambda}$ and $H_{I\lambda}$ in the following manner
\begin{eqnarray}
H_{0\lambda}&=&\sum_{l_{3}m_{3}n_{3}}\hbar\omega_{l_{3}n_{3}\lambda}:b_{l_{3}m_{3}n_{3}}^{\dag}b_{l_{3}m_{3}n_{3}}:+\sum_{m\sigma}\varepsilon_{m\lambda}
:c_{m\sigma}^{\dag}c_{m\sigma}:\\,\label{eq:H0}
H_{I\lambda}&=&\sum_{m_{1}m_{2}}\sum_{l_{3}n_{3}}\sum_{\sigma}M_{m_{1}m_{2}l_{3}n_{3}\lambda}:c_{m_{1}\sigma}^{\dag}c_{m_{2}\sigma}
b^{\dag}_{l_{3}m_{1}-m_{2}n_{3}}:+M^{*}_{m_{1}m_{2}l_{3}n_{3}\lambda}:c_{m_{2}\sigma}^{\dag}c_{m_{1}\sigma}
b_{l_{3}m_{1}-m_{2}n_{3}}:\nonumber\\
& &
+O\left(M^{2}\right),\label{eq:HI}
\end{eqnarray}
where $\overline{n}_{m_{1}\sigma}$ is the average occupancy of an electronic state at the HOS, :: denotes normal ordering, and the following shortened notation was used
\begin{eqnarray*}
& &\varepsilon_{m\lambda}=\epsilon_{lmn\lambda}\\
& &c_{m\sigma}=c_{lmn\sigma}\\
& &M_{m_{1}m_{2}l_{3}n_{3}\lambda}=M^{nnn_{3}}_{lll_{3}m_{1}m_{2}m_{1}-m_{2}\lambda}.
\end{eqnarray*}
The average occupancy is equal to $N_{e}/(4l+2)$ if the HOS is degenerate, where $N_{e}$ is the total number of electrons in the HOS, while $N_{e}$ in the deformed particles is determined by the Fermi-Dirac distribution. The term $O\left(M^{2}\right)$ contains new interactions generated by the transformation, including the effective electron-electron interaction.

The renormalization of the matrix elements of a general Hamiltonian matrix element $H_{ij\lambda}$ is determined through
\begin{equation}
\frac{dH_{ij\lambda}}{d\lambda}=u_{ij\lambda}\left[\eta_{\lambda},H_{I\lambda}\right]_{ij}+r_{ij\lambda}\frac{d\mbox{ln}u_{ij\lambda}}{d\lambda}H_{ij\lambda},
\label{eq:hamiltonian flow}
\end{equation}
and the matrix elements of the generator of the transformation (denoted by $\eta_{ij\lambda}$) is given by
\begin{equation}
\eta_{ij\lambda}=-\frac{r_{ij\lambda}}{E_{i\lambda}-E_{j\lambda}}\left(\left[\eta_{\lambda},H_{I\lambda}\right]_{ij}-\frac{d\mbox{ln}u_{ij}}{d\lambda}
H_{ij\lambda}\right),\label{eq:generator}
\end{equation}
where the states $i$ and $j$ are eigenstates of the diagonal part of the renormalized Hamiltonian $H_{0\lambda}$, with eigenenergies $E_{i\lambda}$ and $E_{j\lambda}$. The properties of the function $u_{ij\lambda}$ are discussed below. The detailed derivation of Eqs.\ \eqref{eq:hamiltonian flow} and \eqref{eq:generator} can be found, for example, in the original paper by Glazek and Wilson \cite{glazek} or in the work of Mielke.\ \cite{mielke1}

The function $u_{ij\lambda}$ is a continuous and differentiable function of the argument $\left(\lambda-\left|E_{i\lambda}-E_{j\lambda}\right|\right)\times\beta$. It is equal to one if $\left(\lambda-\left|E_{i\lambda}-E_{j\lambda}\right|\right)\times\beta<<0$ and to zero if $\left(\lambda-\left|E_{i\lambda}-E_{j\lambda}\right|\right)\times\beta>>0$, where $\beta^{-1}$ is the width in which $u_{ij\lambda}$ varies from one to zero around $\left|E_{i\lambda}-E_{j\lambda}\right|=\lambda$. We use a single value of $\beta$ to all $u_{ij\lambda}$ and take $\beta^{-1}$ to be small so $u_{ij\lambda}$ varies sharply around $\left|E_{i\lambda}-E_{j\lambda}\right|=\lambda$. The meaning of ``small'' $\beta^{-1}$ and ``sharp variation'' is given in what follows.

The function $r_{ij\lambda}$ is defined as $r_{ij\lambda}=1-u_{ij\lambda}$. The factor $r_{ij\lambda}$ multiplying the right-hand-side (RHS) of  Eq.\ \eqref{eq:generator} [and thus also the RHS of Eq.\ \eqref{eq:hamiltonian flow}] ensures that energy denominators small compared to $\lambda$ $\left(E_{i\lambda}-E_{j\lambda}\leq \lambda\right)$ are avoided.

Eqs.\ \eqref{eq:hamiltonian flow} and \eqref{eq:generator} cannot be solved explicitly. Instead, we assume that it is possible to expand the generator $\eta_{\lambda}$ and the transformed Hamiltonian $H_{\lambda}$ in powers of the electron-phonon coupling coefficients $M_{m_{1}m_{2}l_{3}n_{3}\lambda}$ and disregard any terms in $H_{\lambda}$ which are proportional to the third power of $M_{m_{1}m_{2}l_{3}n_{3}\lambda}$ or more.

The lowest order contribution to $\eta_{\lambda}$ comes from the second term in the parentheses in Eq.\ \eqref{eq:generator} and is linear in $M_{m_{1}m_{2}l_{3}l_{3}n_{3}\lambda}$. The factor $r_{ij\lambda}$ ensures that $H_{0\lambda}$ do not contribute to $\eta_{\lambda}$. The renormalization of $H_{\lambda}$ is affected by the generator through the commutator on the RHS of Eq.\ \eqref{eq:hamiltonian flow}. Thus, only the lowest order term of the generator affects the renormalization of the Hamiltonian up to the second order in $M_{m_{1}m_{2}l_{3}n_{3}\lambda}$. We obtain the following expression for $\eta_{\lambda}$
\begin{eqnarray}
\eta_{\lambda}&=&\sum_{m_{1}m_{2}}\sum_{l_{3}n_{3}}\sum_{\sigma}\eta_{m_{1}m_{2}l_{3}n_{3}\lambda}
:c^{\dag}_{m_{1}\sigma}c_{m_{2}\sigma}b^{\dag}_{l_{3}m_{1}-m_{2}n_{3}}:-\eta^{*}_{m_{1}m_{2}l_{3}n_{3}\lambda}
:c^{\dag}_{m_{2}\sigma}c_{m_{1}\sigma}b_{l_{3}n_{3}m_{1}-m_{2}n_{3}}:\nonumber\\
& &+O\left(M^{2}\right),\label{eq:specific_generator}
\end{eqnarray}
where
\begin{equation}
\eta_{m_{1}m_{2}l_{3}n_{3}\lambda}=\frac{-r_{m_{1}m_{2}l_{3}n_{3}\lambda}}{\varepsilon_{m_{1}\lambda}-\varepsilon_{m_{2}\lambda}
+\hbar\omega_{l_{3}n_{3}\lambda}}\frac{d\mbox{ln}u_{m_{1}m_{2}l_{3}n_{3}\lambda}}{d\lambda}
M_{m_{1}m_{2}l_{3}n_{3}}+O\left(M^{3}\right),\label{eq:flow generator}
\end{equation}
and
\begin{equation}
u_{m_{1}m_{2}l_{3}n_{3}\lambda}=u\left[\left(\lambda-\left|\varepsilon_{m_{1}\lambda}-\varepsilon_{m_{2}\lambda}
+\hbar\omega_{l_{3}n_{3}\lambda}\right|\right)\times \beta\right].\label{eq:u}
\end{equation}

We note that we do not take into account corrections to the electron-phonon matrix elements arising from deviations from spherical symmetry. Therefore, the electron-phonon interaction cannot change the energy structure within the HOS (although the inter-shell energy structure is modified) and all electronic energy differences $\varepsilon_{m_{1}\lambda}-\varepsilon_{m_{2}\lambda}$ are in fact $\lambda$-independent. Thus, in the rest of the calculation, as well as in the expression for the renormalization of the generator matrix elements \eqref{eq:flow generator}, we replace $\varepsilon_{m_{1}\lambda}-\varepsilon_{m_{2}\lambda}$ with the initial energy differences $\varepsilon_{m_{1}\Lambda}-\varepsilon_{m_{2}\Lambda}$, where $\varepsilon_{m\Lambda}$ are the non-renormalized energies of the electrons.

Inserting the generator \eqref{eq:specific_generator} into Eq.\ \eqref{eq:hamiltonian flow} we find that the commutator in Eq.\ \eqref{eq:hamiltonian flow} contributes to the renormalization of $M_{m_{1}m_{2}l_{3}n_{3}\lambda}$ only to the third power (or more) of the electron-phonon coupling. Therefore, the renormalization is determined through the following equation
\begin{equation}
\frac{dM_{m_{1}m_{2}l_{3}n_{3}\lambda}}{d\lambda}=r_{m_{1}m_{2}l_{3}n_{3}\lambda}\frac{d\mbox{ln}u_{m_{1}m_{2}l_{3}n_{3}\lambda}}{d\lambda}
M_{m_{1}m_{2}l_{3}n_{3}\lambda}+O\left(M^{3}\right),\label{eq:flow coupling}
\end{equation}
with the solution
\begin{equation}
M_{m_{1}m_{2}l_{3}n_{3}\lambda}=M_{m_{1}m_{2}l_{3}n_{3}\Lambda}e_{m_{1}m_{2}l_{3}n_{3}\lambda}
=M_{m_{1}m_{2}l_{3}n_{3}\Lambda}u_{m_{1}m_{2}l_{3}n_{3}\lambda}e^{r_{m_{1}m_{2}l_{3}n_{3}\lambda}}.\label{eq:coupling}
\end{equation}
Note that the function $e_{m_{1}m_{2}l_{3}n_{3}\lambda}$ has the same asymptotic behavior as $u_{m_{1}m_{2}l_{3}n_{3}\lambda}$.

The renormalization of the single-particle electron and phonon energies is determined solely by the commutator in Eq.\ \eqref{eq:hamiltonian flow}, while the second term in (\ref{eq:hamiltonian flow}) is irrelevant due to the presence of the pre-factor $r_{m_{1}m_{2}l_{3}n_{3}\lambda}$. By contrast, both terms contribute to the generated electron-electron interaction. We calculate the commutator and use the expressions for the matrix elements of the generator \eqref{eq:flow generator} and the renormalized electron-phonon coupling coefficients \eqref{eq:coupling}, in order to derive differential equations describing the renormalization of the single-particle energies $\varepsilon_{M\Sigma\lambda}$ and $\omega_{l_{3}n_{3}\lambda}$, and the effective interaction $V_{MM^{'}\lambda}$ between time-reversed electron pairs ($\{M\uparrow,-M\downarrow\}$ and $\{M^{'}\uparrow,-M^{'}\downarrow\}$). We obtain the following three differential equations
\begin{eqnarray}
\frac{dV_{MM^{'}\lambda}}{d\lambda}&=&u_{MM^{'}\lambda}\sum_{l_{3}n_{3}}\left(\frac{2}{\varepsilon_{M\Lambda}-\varepsilon_{M^{'}\Lambda}+\hbar\omega_{l_{3}n{3}\lambda}}
e_{M^{'}Ml_{3}n_{3}\lambda}\frac{de_{MM^{'}l_{3}n_{3}\lambda}}{d\lambda}\left|M_{MM^{'}l_{3}n_{3}\Lambda}\right|^{2}\right.\nonumber\\
& &+\left.\frac{2}{\varepsilon_{M^{'}\Lambda}-\varepsilon_{M\Lambda}+\hbar\omega_{l_{3}n{3}\lambda}}
e_{MM^{'}l_{3}n_{3}\lambda}\frac{de_{M^{'}Ml_{3}n_{3}\lambda}}{d\lambda}\left|M_{MM^{'}l_{3}n_{3}\Lambda}\right|^{2}\right)\nonumber\\
& &+\frac{r_{MM^{'}\lambda}}{u_{MM^{'}\lambda}}\frac{du_{MM^{'}\lambda}}{d\lambda}V_{MM^{'}\lambda},\label{eq:flow effective}\\
\frac{d\varepsilon_{M\Sigma\lambda}}{d\lambda}&=& \sum_{ml_{3}n_{3}\sigma}\left(\frac{1-\overline{n}_{m\sigma}}{\varepsilon_{m\sigma\Lambda}-\varepsilon_{M\Sigma\Lambda}+\hbar\omega_{l_{3}n{3}\lambda}}
\left|M_{mMl_{3}n_{3}\Lambda}\right|^{2}\frac{de^{2}_{mMl_{3}n_{3}\lambda}}{d\lambda}\right.\nonumber\\
& &\left.-\frac{\overline{n}_{m\sigma}}{\varepsilon_{M\Sigma\Lambda}-\varepsilon_{m\sigma\Lambda}+\hbar\omega_{l_{3}n{3}\lambda}}
\left|M_{Mml_{3}n_{3}\Lambda}\right|^{2}\frac{de^{2}_{Mml_{3}n_{3}\lambda}}{d\lambda}\right),\label{eq:derivative epsilon}\\
\frac{d\omega_{l_{3}n_{3}\lambda}}{d\lambda}=& &\sum_{l_{3}n_{3}m_{1}\sigma}\frac{\overline{n}_{m_{1}\sigma}-\overline{n}_{m_{1}-m_{3}\sigma}}
{\varepsilon_{m_{1}\sigma\Lambda}-\varepsilon_{m_{1}-m_{3}\sigma\Lambda}+\hbar\omega_{l_{3}n_{3}\lambda}}
\left|M_{m_{1}m_{1}-m_{3}l_{3}n_{3}\Lambda}\right|^{2}\frac{de^{2}_{m_{1}m_{1}-m_{3}l_{3}n_{3}\lambda}}{d\lambda},\label{eq:omega flow}
\end{eqnarray}
where
\begin{equation}
u_{MM^{'}\lambda}=u\left[\left(\lambda-2\left|\varepsilon_{M\sigma\Lambda}-\varepsilon_{M^{'}\sigma\Lambda}\right|\right)\times \beta\right].\label{eq:uMMp}
\end{equation}

From Eq.\ \eqref{eq:omega flow} we see that the renormalization of the phonon frequencies is zero (at least up to the second order in $M_{m_{1}m_{2}l_{3}n_{3}\lambda}$) when we consider spherical nanoparticles with degenerate energy shells. Even if deviations from spherical symmetry are taken into account, the corrections are small compared to the non-renormalized frequencies. Therefore, we neglect the renormalization of the phonon frequencies and assume that they are $\lambda$-independent.

The solution of \eqref{eq:flow effective} is given by
\begin{equation}
V_{MM^{'}\lambda_{min}}=-\mbox{exp}\left(\int^{\Lambda}_{\lambda_{min}}p(t)dt\right)\int^{\Lambda}_{\lambda_{min}}
\mbox{exp}\left(-\int^{\Lambda}_{s}p(t)dt\right)g(s)ds,\label{eq:solution V}
\end{equation}
where we used the fact that $V_{MM^{'}\Lambda}=0$, and the functions $p(\lambda)$ and $g(\lambda)$ are defined as
\begin{eqnarray}
g(\lambda)&=&u_{MM^{'}\lambda}\left(\sum_{l_{3}n_{3}}\frac{2}{\varepsilon_{M\sigma\Lambda}-\varepsilon_{M^{'}\sigma\Lambda}+\hbar\omega_{l_{3}n{3}\lambda}}
e_{M^{'}Ml_{3}n_{3}\lambda}\frac{de_{MM^{'}l_{3}n_{3}\lambda}}{d\lambda}\left|M_{MM^{'}l_{3}n_{3}\Lambda}\right|^{2}\right.\nonumber\\
&&+\sum_{l_{3}n_{3}}\left.\frac{2}{\varepsilon_{M^{'}\sigma\Lambda}-\varepsilon_{M\sigma\Lambda}+\hbar\omega_{l_{3}n{3}\lambda}}
e_{MM^{'}l_{3}n_{3}\lambda}\frac{de_{M^{'}Ml_{3}n_{3}\lambda}}{d\lambda}\left|M_{MM^{'}l_{3}n_{3}\Lambda}\right|^{2}\right)\nonumber\\
&=&u_{MM^{'}\lambda}\left(\sum_{l_{3}n_{3}}g^{1}_{l_{3}n_{3}}+\sum_{l_{3}n_{3}}g^{2}_{l_{3}n_{3}}\right),\label{eq:glambda}\\
p(\lambda)&=&-\frac{r_{MM^{'}\lambda}}{u_{MM^{'}\lambda}}\frac{du_{MM^{'}\lambda}}{d\lambda}.\label{eq:plambda}
\end{eqnarray}

In order to explicitly calculate $V_{MM^{'}\lambda_{min}}$ we need to choose a specific form of $u_{MM^{'}l_{3}n_{3}\lambda}$ and $u_{MM^{'}\lambda}$. We define $r_{MM^{'}l_{3}n_{3}\lambda}$ and $r_{MM^{'}\lambda}$ to be Fermi-Dirac functions with a width that we denoted before as $\beta^{-1}$. These functions (as well as $u_{MM^{'}l_{3}n_{3}\lambda}$ and $u_{MM^{'}\lambda}$) vary sharply from zero to one around $\lambda=\lambda_{MM^{'}l_{3}n_{3}}$ and $\lambda=\lambda_{MM^{'}}$ respectively, where $\lambda_{MM^{'}l_{3}n_{3}}$ and $\lambda_{MM^{'}}$ are defined as
\begin{eqnarray}
\lambda_{MM^{'}l_{3}n_{3}}&=&\left|\varepsilon_{M\sigma\Lambda}-\varepsilon_{M^{'}\sigma\Lambda}+\hbar\omega_{l_{3}n{3}}\right|,\label{eq:lambdap}\\ \lambda_{MM^{'}}&=&2\left|\varepsilon_{M\sigma\Lambda}-\varepsilon_{M^{'}\sigma\Lambda}\right|.\label{eq:lambdappp}
\end{eqnarray}

Each term in the two sums in \eqref{eq:glambda} is characterized by certain $\lambda_{MM^{'}l_{3}n_{3}}$ and  $\lambda_{M^{'}Ml_{3}n_{3}}$ at which the functions $e_{MM^{'}l_{3}n_{3}\lambda}$ and $e_{M^{'}Ml_{3}n_{3}\lambda}$ vary from one to zero as $\lambda$ is lowered. These variations result in  formation of peaks in $g\left(\lambda\right)$ around the various $\lambda_{MM^{'}l_{3}n_{3}}$ and  $\lambda_{M^{'}Ml_{3}n_{3}}$. We take $\beta^{-1}$ to be much smaller than any energy difference in our system, and therefore ensure that the variation from one to zero of $e_{MM^{'}l_{3}n_{3}\lambda}$ and $e_{M^{'}Ml_{3}n_{3}\lambda}$ is fast enough, so there is only a small overlap between the various peaks in $g\left(\lambda\right)$. The smallness of $\beta^{-1}$ also ensures that there is only a small overlap between the peak located around the lowest value of $\lambda_{MM^{'}l_{3}n_{3}}$ and the drop (from one to zero) in $p(\lambda)$ at $\lambda_{MM^{'}}$.

If effects of deformations on the electron-phonon matrix elements are taken into account, then the difference $\varepsilon_{m_{1}\lambda}-\varepsilon_{m_{2}\lambda}$ becomes $\lambda$-dependent. In this case we need to impose an additional condition on the size of $\beta^{-1}$. We choose  $\beta^{-1}$ to be small enough so the change in   $\left|\varepsilon_{M\sigma\lambda}-\varepsilon_{M^{'}\sigma\lambda}\right|$, as $\lambda$ is varied across $\lambda_{MM^{'}l_{3}n_{3}}$ or $\lambda_{MM^{'}}$, is small compared to either $\lambda_{MM^{'}l_{3}n_{3}}$ or $\lambda_{MM^{'}}$. This last condition can be written as
\begin{equation}\label{eq:condition beta}
\beta^{-1}\ll\left|\left(\left.\frac{d\left|\varepsilon_{M\sigma\lambda}-\varepsilon_{M^{'}\sigma\lambda}\right|}{d\lambda}
\right|_{\lambda=\lambda_{MM^{'}l_{3}n_{3}},\lambda_{MM^{'}}}\right)^{-1}\times\left(\lambda=\lambda_{MM^{'}l_{3}n_{3}},\lambda_{MM^{'}}\right)\right|,
\end{equation}
for all $\lambda_{MM^{'}}$ and $\lambda_{MM^{'}l_{3}n_{3}}$.

Using the specific form we chose for $u_{M^{'}Ml_{3}n_{3}\lambda}$, we can show that  $e_{M^{'}Ml_{3}n_{3}\lambda}$ is essentially equal to $u_{M^{'}Ml_{3}n_{3}\lambda}$ even in the transition zone around $\lambda_{MM^{'}l_{3}n_{3}}$ (and not just as $\lambda$ tends to zero or to $\Lambda$). We can also show that the terms in the first sum of $g(\lambda)$ \eqref{eq:glambda} with $\lambda_{MM^{'}l_{3}n_{3}}$ larger than both $\lambda_{M^{'}Ml_{3}n_{3}}$ and $\lambda_{MM^{'}}$ give a finite non-zero contribution to $V_{MM^{'}\lambda_{min}}$, as long as $\lambda_{min}$ is smaller than $\lambda_{MM^{'}l_{3}n_{3}}$ but larger than $\lambda_{MM^{'}}$. Furthermore, due to the narrowness of the variation in $g^{1}_{l_{3}n_{3}}\left(\lambda\right)$ around $\lambda_{MM^{'}l_{3}n_{3}}$ the integral $\int^{\Lambda}_{\lambda_{min}}\mbox{exp}\left(-\int^{\Lambda}_{s}p(t)dt\right)g^{1}_{l_{3}n_{3}}(s)ds$ is independent of $\lambda_{min}$, as long as $\lambda_{min}$ is slightly smaller than $\lambda_{MM^{'}l_{3}n_{3}}$.

On the other hand, all terms in the first sum of $g(\lambda)$ with $\lambda_{MM^{'}l_{3}n_{3}}$ smaller than either $\lambda_{M^{'}Ml_{3}n_{3}}$ or $\lambda_{MM^{'}}$ give exponentially small contributions, regardless of the exact value of $\lambda_{min}$. The same is true for the second sum in \eqref{eq:glambda} with the roles of $\lambda_{MM^{'}l_{3}n_{3}}$ and $\lambda_{M^{'}Ml_{3}n_{3}}$ interchanged.

As long as $\lambda_{min}$ is larger than $\lambda_{MM^{'}}$ the exponential pre-factor in \eqref{eq:solution V} is approximately equal to one. However, if $\lambda_{min}$ is chosen to be smaller than $\lambda_{MM^{'}}$ then the pre-factor becomes $\textrm{exp}\left(-\lambda_{MM^{'}}\beta\right)$, and the effective interaction is suppressed. This is in accordance with the general scheme of the similarity renormalization, since $\lambda_{MM^{'}}$ is just the difference between the single-electron energies of the pair $\{M\uparrow,-M\downarrow\}$ and the pair $\{M^{'}\uparrow,-M^{'}\downarrow\}$.

Therefore, if we try to completely diagonalize the Hamiltonian (up to the second order in $M_{m_{1}m_{2}l_{3}n_{3}\lambda}$) by taking $\lambda_{min}$ to zero, we obtain decoupled free electrons and phonons with renormalized energies. This is not surprising since our application of the similarity renormalization method is perturbative in its nature, and one cannot hope to obtain the paired ground-state of the electrons from the single-electron ground-state via a perturbation expansion. However, our aim is not to completely diagonalize the Hamiltonian but to decouple the electrons from the phonons sufficiently so as to obtain the effective electron-electron interaction while accounting for all relevant mediating phonons. This goal is achieved by carrying the integration in Eq.\ \eqref{eq:solution V} down to $\lambda_{min}$ which is slightly smaller than the smallest of $\lambda_{MM^{'}l_{3}n_{3}}$ or $\lambda_{M^{'}Ml_{3}n_{3}}$ that is still larger than $\lambda_{MM^{'}}$.

The contribution from the first sum in $g(\lambda)$ \eqref{eq:glambda} to $V_{MM^{'}\lambda_{min}}$ (as long as $\lambda_{min}$ is larger than $\lambda_{MM^{'}}$ and smaller than the smallest relevant $\lambda_{MM^{'}l_{3}n_{3}}$ and $\lambda_{M^{'}Ml_{3}n_{3}}$) is
\begin{equation}
V^{\textrm{first sum}}_{MM^{'}\lambda_{min}}=\sum_{l_{3}n_{3}}\frac{-2\left|M_{MM^{'}l_{3}n_{3}\Lambda}\right|^{2}}
{\varepsilon_{M\Lambda}-\varepsilon_{M^{'}\Lambda}+\hbar\omega_{l_{3}n_{3}}},\label{eq:first sum}
\end{equation}
where $\varepsilon_{M\Lambda}>\varepsilon_{M^{'}\Lambda}$ and $\varepsilon_{M\Lambda}-\varepsilon_{M^{'}\Lambda}<\hbar\omega_{l_{3}n_{3}}$. The contribution from the second sum is
\begin{equation}
V^{\textrm{second sum}}_{MM^{'}\lambda_{min}}=\sum_{l_{3}n_{3}}\frac{-2\left|M_{MM^{'}l_{3}n_{3}\Lambda}\right|^{2}}
{\varepsilon_{M^{'}\Lambda}-\varepsilon_{M\Lambda}+\hbar\omega_{l_{3}n_{3}}},\label{eq:seond sum}
\end{equation}
where $\varepsilon_{M^{'}\Lambda}>\varepsilon_{M\Lambda}$ and $\varepsilon_{M^{'}\Lambda}-\varepsilon_{M\Lambda}<\hbar\omega_{l_{3}n_{3}}$. Both contributions do not depend explicitly on $\lambda_{min}$. We see that either the first or the second term contribute to the interaction matrix element, depending on the sign of $\varepsilon_{M\Lambda}-\varepsilon_{M^{'}\Lambda}$. Therefore, we can write the interaction matrix element as
\begin{equation}
V_{MM^{'}}=\sum_{l_{3}n_{3}}\frac{-2\left|M_{MM^{'}l_{3}n_{3}\Lambda}\right|^{2}}
{\left|\varepsilon_{M\Lambda}-\varepsilon_{M^{'}\Lambda}\right|+\hbar\omega_{l_{3}n_{3}}}
\Theta\left(\hbar\omega_{l_{3}n_{3}}-\left|\varepsilon_{M\Lambda}-\varepsilon_{M^{'}\Lambda}\right|\right).
\label{eq:interactionfull}
\end{equation}

Inspecting the expression in \eqref{eq:interactionfull} we can see that the resulting effective interaction is always attractive and decays monotonically with the increasing separation between the electronic energy levels. Furthermore, a phonon cannot couple electron pairs with energies that are separated by more than its energy. The result in \eqref{eq:interactionfull} coincides with the effective interaction obtained by the application of the Fr\"{o}hlich transformation \eqref{eq:one shell hamiltonian}, if the nanoparticle is completely spherical. We note that if inter-shell interaction is not neglected then Eq.\ \eqref{eq:interactionfull} includes a summation over all relevant shells as well as inter-shell terms. In this case the non-renormalized energies of the electrons in Eq.\ \eqref{eq:interactionfull} are replaced by the $\lambda$-dependent renormalized energies.

The renormalization of the single-particle electron energies is obtained by integrating Eq.\ \eqref{eq:derivative epsilon}. The integration from $\lambda_{min}$ up to $\Lambda$ yields
\begin{eqnarray}
\varepsilon_{M\Sigma\lambda_{min}}&=&\varepsilon_{M\Sigma\Lambda}+\sum\mbox{$^{^{'}}$}\frac{\left(\overline{n}_{m\sigma}-1\right)
\left|M_{mMl_{3}n_{3}\Lambda}\right|^{2}}{\varepsilon_{m\sigma\Lambda}-\varepsilon_{M\Sigma\Lambda}+\hbar\omega_{l_{3}n{3}}}\nonumber\\
&&+\sum\mbox{$^{^{''}}$}\frac{\overline{n}_{m\sigma}\left|M_{Mml_{3}n_{3}\Lambda}\right|^{2}}{\varepsilon_{M\Sigma\Lambda}-\varepsilon_{m\sigma\Lambda}
+\hbar\omega_{l_{3}n{3}}}.\label{eq:epsilon M}
\end{eqnarray}
The sum $\sum^{'}$ in \eqref{eq:epsilon M} runs over all values of $\varepsilon_{m\sigma\Lambda}$ and $\hbar\omega_{l_{3}n{3}}$ for which    $\lambda_{mMl_{3}n_{3}}$ is larger than $\lambda_{min}$, and the sum $\sum^{''}$ runs over all values of $\varepsilon_{m\sigma\Lambda}$ and $\hbar\omega_{l_{3}n{3}}$ for which $\lambda_{Mml_{3}n_{3}}$ is larger than $\lambda_{min}$. In order to take into account the effect of all relevant phonons we need to take $\lambda_{min}$ to be smaller than the energy of the least energetic spheroidal phonon that can interact with electrons in the HOS.

We now consider an initial untransformed Hamiltonian that includes a Coulomb interaction term. In principal, the additional Coulomb term should modify the generator and the flow of the Hamiltonian. However, if we retain the generator \eqref{eq:specific_generator} and do not include in it terms arising from Coulomb interaction, then the flow of the electron-electron effective interaction remains unmodified up to the second order in the interaction coefficients. Therefore, the only modification introduced into the calculation is the appearance of the term $M_{MM^{'}}^{co}$ on the RHS of Eq.\ \eqref{eq:interactionfull}, since the electron-electron interaction is equal to the Coulomb term and not to zero at $\lambda=\Lambda$. Furthermore, the inclusion of the Coulomb term in $H_{I\lambda}$ does not affect the renormalization of the electron energies as long as the generator \eqref{eq:specific_generator} is retained, since the commutator on RHS of \eqref{eq:hamiltonian flow} do not yield additional terms that contribute to the renormalization of the electron energies up to the second order (including) in the electron-phonon and Coulomb interaction coefficients. The generator \eqref{eq:specific_generator} cannot induce the elimination of the Coulomb interaction, which even after the transformation is still able to couple states with large energy difference.

We note that Mielke \cite{mielke1} solved the equivalent Eq.\ to \eqref{eq:flow effective} in the bulk while assuming $u_{kk^{'}q\lambda}=1$, which is equivalent to assuming $u_{MM^{'}\lambda}=1$ in our calculation. This assumption leads to $p(\lambda)=0$ and to the disappearance of the factor $u_{MM^{'}\lambda}$, which multiplies the sum in \eqref{eq:glambda}. This last factor is responsible for the Heaviside function in \eqref{eq:interactionfull}. Therefore, in Mielke's interaction, phonons with energy smaller than $\left|\varepsilon_{M\Lambda}-\varepsilon_{M^{'}\Lambda}\right|$ can contribute to the matrix element $V_{MM^{'}}$ and therefore artificially enhance the effective interaction. Furthermore, the effective interaction of Mielke is not eliminated in the limit $\lambda\rightarrow 0$. This result is inconsistent with the general scheme of the similarity renormalization method.

It should be noted that Mielke investigated a model in which the effective interaction was mediated by non-dispersive phonons with an energy that is much larger than the energy of the electrons. In this model, and as long as $\lambda_{min}$ is taken to be slightly smaller than the single frequency of the non-dispersive phonons, our results and the results of Mielke coincide.


\begin{acknowledgments}
This research was supported by the German-Israeli Foundation (GIF) through Grant No. 981-185.14/2007
\end{acknowledgments}

\end{document}